\documentclass[12pt,a4paper]{article}
\usepackage{amsmath}
\usepackage{amssymb}
\usepackage{graphicx}
\usepackage{url}
\usepackage{wrapfig}
\usepackage{multicol}
\usepackage{multirow}
\usepackage{authblk}
\usepackage{esint}
\usepackage{floatflt}
\usepackage[font=small,labelfont=bf]{caption}
\usepackage{subcaption}
\usepackage{mathtools}
\usepackage{sectsty}
\usepackage{hyperref}
\usepackage{appendix}
\usepackage{algorithm2e}
\usepackage{cite}
\hypersetup{
    colorlinks=true,
    allcolors=blue,
}

\newcommand{\eq}[2]{\begin{equation} \label{#1} #2 \end{equation}}
\newcommand{\meq}[2]{\begin{equation}\begin{aligned} \label{#1} #2 \end{aligned}\end{equation}}

\title{\bf Reaction-diffusion kinetics on lattice at the microscopic scale}
\date{}

\author{\small Wei-Xiang Chew$^{1,2}$, Kazunari Kaizu$^{1}$, Masaki Watabe$^{1}$, Sithi V. Muniandy$^{2}$, Koichi Takahashi$^{1}$ and Satya N. V. Arjunan$^{1*}$}
\affil{\small
$^{1}$ Laboratory for Biologically Inspired Computing, RIKEN Center for Biosystems Dynamics Research, Suita, Osaka, Japan\\
$^{2}$ Department of Physics, Faculty of Science, University of Malaya, 50603, Kuala Lumpur, Malaysia.\\
$^{*}$E-mail: satya@riken.jp 
}

\begin{document}
\maketitle
\sectionfont{\large\MakeUppercase}
\section*{Abstract}
Lattice-based stochastic simulators are commonly used to study biological reaction-diffusion processes. Some of these schemes that are based on the reaction-diffusion master equation (RDME), can simulate for extended spatial and temporal scales but cannot directly account for the microscopic effects in the cell such as volume exclusion and diffusion-influenced reactions. Nonetheless, schemes based on the high-resolution microscopic lattice method (MLM) can directly simulate these effects by representing each finite-sized molecule explicitly as a random walker on fine lattice voxels. The theory and consistency of MLM in simulating diffusion-influenced reactions have not been clarified in detail. Here, we examine MLM in solving diffusion-influenced reactions in 3D space by employing the Spatiocyte simulation scheme. Applying the random walk theory, we construct the general theoretical framework underlying the method and obtain analytical expressions for the total rebinding probability and the effective reaction rate. By matching Collins-Kimball and lattice-based rate constants, we obtained the exact expressions to determine the reaction acceptance probability and voxel size. We found that the size of voxel should be about $2\%$ larger than the molecule. The theoretical framework of MLM is validated by numerical simulations, showing good agreement with the off-lattice particle-based method, eGFRD. MLM run time is more than an order of magnitude faster than eGFRD when diffusing macromolecules with typical concentrations observed in the cell. MLM also showed good agreements with eGFRD and mean-field models in case studies of two basic motifs of intracellular signaling, the protein production-degradation process and the dual phosphorylation-dephosphorylation cycle. In addition, when a reaction compartment is populated with volume-excluding obstacles, MLM captures the non-classical reaction kinetics caused by anomalous diffusion of reacting molecules.

\newpage
\section{Introduction}
\paragraph{}
In the intracellular environment, macromolecules can be heterogeneously distributed in space and react stochastically at low concentrations. The conventional mass action-based approach is insufficient to describe the reaction-diffusion (RD) behavior of the macromolecules and thus, it is necessary to incorporate space and stochasticity into the model \cite{howard2003,klann2012,engblom17,william2018,earnest2018,smith2018spatial}. Generally, we can represent space as a continuum (off-lattice) or a discretized lattice model. In the former, each molecule is represented as a point or a hard-body sphere that propagates via Brownian motion in continuous space \cite{stiles1998,plimpton2003,van2005,Sanford2006,ridgway2008,Byrne2010,gruenert2010,takahashi2010,
klann2012b,schoneberg2013,andrews2016,michalski2016,lehnert2017,mlspace,bashardanesh2017,
donev2018,sayyidmousavi2018reactive} 
. Bimolecular reaction is often modeled as a collision-based interaction \cite{stiles1998,ridgway2008,Byrne2010,takahashi2010,lehnert2017,bashardanesh2017} according to the Smoluchowski model of diffusion-influenced reactions \cite{smoluchowski1917,Collins1949}. In some models, the finite size of molecules is taken into account in the reaction and therefore, the effects of volume-exclusion by molecules can be reproduced \cite{ridgway2008,Byrne2010,takahashi2010,klann2012b,gruenert2010,schoneberg2013,michalski2016,andrews2016,bashardanesh2017}. Although continuous space-time models are physically consistent, the cost of computation becomes significant when simulating non-dilute and crowded conditions in the cell \cite{Takahashi2005}. 

On the other hand, in lattice approaches, the average diffusion behavior is adopted and the reactions follow either the simple first-order process, or the second-order process when two reactive molecules meet on the same lattice voxel. Such approaches reduce the computational cost even in crowded space and provide an efficient way to simulate large numbers of molecules and reactions. 
Within lattice approaches, variation exists depending on how each molecule is represented and reaction is modeled. In the well-established reaction-diffusion master equation (RDME) models \cite{baras1996,hattne2005,drawert2012,cowan2012,hepburn12,roberts2013Lattice,drawert2016}, space is discretized into lattice voxels called subvolumes. In each subvolume, point-like molecules are assumed to be dilute and well-mixed. To obey the well-mixed condition, there is a limit to the size of the subvolume \cite{Hellander2012,hattne2005,hellander2015}, which in turn imposes a limit to the spatial resolution. Diffusion of molecules across subvolumes is modeled as a first-order reaction with a concentration dependent rate. Unimolecular and bimolecular reactions only occur within each subvolume with a rate defined by the propensity function \cite{gillespie1976}. Compared to continuum-based schemes, RDME models RD from the mesoscopic to the macroscopic scale but not at the microscopic scale. However, there have been several efforts to overcome the well-mixed limit in RDME models and to bridge mesoscopic and microscopic scales \cite{Fange2010,Hellander2012,hellander2015,isaacson2013}. 

Apart from the RDME lattice models, there is another class of schemes, which we refer as microscopic lattice method (MLM) that represents molecules at single particle resolution \cite{Torney1983,montalenti2000,Saxton2002,Berry2002a,Schnell2004,Grima2006,Boulianne2008a,
schmit2009,Arjunan2010,Gillespie2014,pitulice2014,gomez2015}. In most of these schemes \cite{Torney1983,Saxton2002,Berry2002a,Schnell2004,Grima2006,Boulianne2008a,Arjunan2010,gomez2015}, the size of the voxel follows the molecule size, whereas in the small-voxel tracking algorithm (SVTA) \cite{Gillespie2014}, a particle can occupy multiple voxels, providing greater spatial resolution at the cost of higher computational complexity. In MLM, a molecule hops into a neighbor voxel at a constant rate such that normal diffusion is satisfied. Excluded volume arises naturally since the size of molecule is directly reflected by the voxel size and occupancy in the lattice. Similar to RDME models, unimolecular reaction is modeled as a first-order process. Bimolecular reactions are coupled to molecular collisions in all of these schemes except GridCell \cite{Boulianne2008a}. In the collision-based reaction schemes, the steady-state reaction rate follows the macroscopic effective reaction rate when the reaction is activation-limited. However, the reaction accuracy of MLM has not been studied in detail when it is diffusion-influenced. In a recent work, Sturrock \cite{Sturrock2016} also reported several shortcomings in MLM, notably in the accuracy of Spatiocyte \cite{Arjunan2010} when estimating steady-state bimolecular reaction rates.

Our focus in this work is to examine in detail the accuracy and consistency of MLM in solving diffusion-influenced reactions using theoretical analysis and numerical simulations. The theoretical framework here is constructed based on the hexagonal close-packed (HCP) lattice but is also applicable for any regular lattice arrangements such as the simple cubic lattice. We employ the Spatiocyte scheme to construct and analyze the general theoretical framework of MLM in both activation- and diffusion-limited regimes. We then describe the first-passage behavior of the method according to the random walk theory and obtain the analytical formula for the total rebinding probability of a pair of reacting molecules and their effective reaction rate constant. Next, we perform numerical simulations to evaluate the accuracy of the theory and investigate the time-dependent kinetics. We found that MLM exhibits the expected steady-state and asymptotic time-dependent behaviors of the reaction as in the collision-based continuum model. Subsequently, we evaluate the performance of MLM in comparison to other well-known off-lattice methods. As application examples, we show that the method correctly recapitulates the time-dependent behavior of proteins in the production-degradation process and the dual phosphorylation-dephosphorylation cycle, two fundamental building blocks of intracellular signaling. Finally, we demonstrate the effects of crowding obstacles on the kinetics of a simple bimolecular reaction with MLM.

\section{Methods}
We begin by presenting the theoretical background of the Collins-Kimball \cite{Collins1949} approach in modeling irreversible bimolecular diffusion-influenced reaction. We highlight the particle-pair formalism for the reaction rate, which will be used in the MLM theory. We then briefly describe the Spatiocyte RD scheme, an MLM implemented on the HCP lattice. Finally, we construct the theoretical framework of reaction rate coefficient on lattice using the Spatiocyte scheme.

\subsection{Irreversible bimolecular diffusion-influenced reaction in continuum-based framework}
Consider an irreversible bimolecular reaction involving two distinct species: 
\eq{birxn}{A+B \xrightarrow{} C,}
where $A$ and $B$ are hard-sphere molecules with radii $r_A$ and $r_B$, respectively. The molecules diffuse in three-dimensional (3D) space with diffusion coefficients $D_A$ and $D_B$. 
The time evolution of the species concentration is well-described by a time-dependent rate coefficient $k_{irr}(t)$:
\eq{}{\frac{d[A](t)}{dt}=\frac{d[B](t)}{dt}=-k_{irr}(t)[A](t)[B](t).} 
Smoluchowski \cite{smoluchowski1917} derived the rate coefficient by relating the diffusion coefficient of the molecules with molecular collisions, leading to the product formation. In his work, $A$ is made up of an immobile molecule and is surrounded by multiple diffusing $B$ molecules. Collins and Kimball extended the Smoluchowski theory by modeling the reaction using radiation boundary condition and obtained the rate coefficient in 3D space as a function of microscopic parameters, namely the microscopic or the intrinsic reaction rate constant, $k_a$, the contact distance of the reacting molecules, $R=r_A+r_B$ and the relative diffusion coefficient of the pair, $D=D_A+D_B$:
\eq{ktck}{
k_{irr}(t) = \frac{k_Dk_a}{k_D+k_a}\left[ 1+\frac{k_a}{k_D}\Phi\left(\frac{k_a}{k_D}\sqrt{\frac{t}{\tau'}}\right)  \right].
} 
Here, $k_D=4\pi RD$ is the collision rate, $\Phi(x)=\exp({x^2})\text{erfc}(x)$ and $\tau'=(1/D)(k_aR/(k_a+k_D))^2$. 
The rate coefficient \eqref{ktck} starts ($t=0$) at $k_a$ but decays rapidly to \cite{agmon1990}
\eq{klarget}{k_{irr}(t)\simeq k_{eff}\left[1+\frac{k_a}{k_a+k_D}\frac{R}{\sqrt{\pi Dt}}\right],}
at long-time. $k_{eff}$ is the steady-state or the effective reaction rate constant given by \cite{noyes1961}: 
\eq{keff1}{
k_{eff}\coloneqq k_{irr}(t\to \infty)=\frac{k_ak_D}{k_a+k_D}.}
\\According to Noyes theory \cite{Noyes1954,berg1978,naqvi1980}, the rate coefficient can be expressed equivalently using the particle-pair approach:
\eq{ppa}{k_{irr}(t) = k_a S(t;R),}
where $S(t;R)$ denotes the survival probability of an isolated reactant pair at time $t$ given that they were initially in contact. Additionally, let $p_{reb}(R,t|R,0)$ denote the rebinding-time probability distribution for a reactive particle-pair separated by distance $R$ at time $t$, given that the pair were initially in contact. In the case of radiation boundary condition, the probability distribution is given by (see Appendix \ref{arpd})
\eq{prebt}{
p_{reb}(R,t|R,0)=\left(\frac{k_a}{4\pi R^3}\right) \left(\frac{k_a}{k_D}+1 \right)\left(\frac{1}{\sqrt{\pi\tau}}-\exp(\tau)\text{erfc}(\sqrt{\tau}) \right)
,}
with $\tau=tD(1+k_a/k_D)^2/{R^2}$. 
\\Note that the survival probability $S(t;R)$ is the same as the probability that the first rebinding event between an initially in-contact pair has not yet occurred at time $t$. Hence we can rewrite Eq. \eqref{ppa} as
\eq{ktnoyes}{k_{irr}(t) = k_a\left[1- \int_0^t  p_{reb}(R,\tau|R,0)d\tau\right].} 
At long-time, we then have
\meq{longtimek}{
k_{eff}\coloneqq k_{irr}(t\to \infty)&=k_a\left[1- \int_0^\infty  p_{reb}(R,\tau|R,0)d\tau\right],
}
where the integrated term gives the total rebinding probability:
\eq{ptot}{
P_{reb}=\int_0^\infty p_{reb}(R,t|R,0)dt=\frac{1}{1+\frac{k_D}{k_a}}.
}
Therefore, the effective rate constant \eqref{keff1} can also be written in terms of the total rebinding probability:
\eq{keff2}{
k_{eff}=k_a(1-P_{reb}).
}
The above relation was also described previously, but in the context of irreversible and reversible rate constants \cite{andrews2004}.
In subsequent sections, we use the relations described by Eqs. \eqref{ktnoyes} and \eqref{keff2} as the central concepts to derive the rate coefficient in MLM.

\begin{figure}[h]   
\centering
  \includegraphics[width=12.0cm]{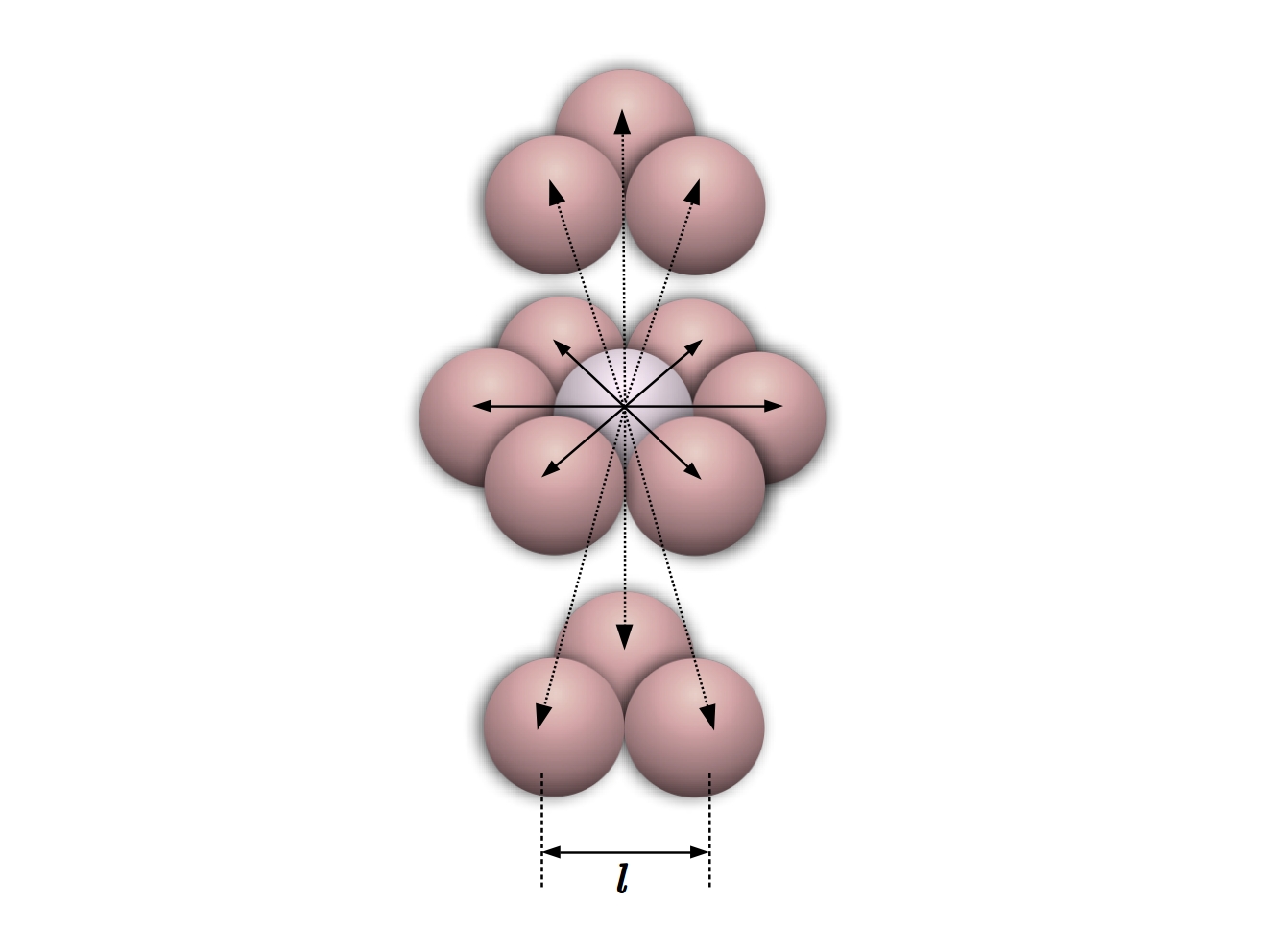}    
  \vspace{0.4cm}
  \caption{ A voxel on the HCP lattice has 12 nearest neighbor voxels. The distance between the centers of two adjacent voxels is the voxel size $l$.
  }
\label{HCP}
\end{figure}

\subsection{Spatiocyte reaction-diffusion scheme}\label{scheme}
In the Spatiocyte scheme (see Algorithm \ref{code}) \cite{Arjunan2010}, space is discretized into HCP lattice because the arrangement allows the highest density of regular sphere voxels in 3D space. The voxel has a diameter $l$ and can be occupied by at most, a single molecule. At each diffusion time step $t_d$, a molecule can hop to one of its 12 nearest neighbor voxels (see Figure \ref{HCP}) with the step acceptance probability $P_w=1$. $t_d=l^2/6D_x$, where $D_x$ is the diffusion coefficient of molecule $x$. Given the irreversible bimolecular reaction in Eq. \eqref{birxn}, a collision arises when $B$ meets $A$ at the destination voxel. The collision is reactive with an acceptance probability $P_a=\Delta N_C/Z$. Here $\Delta N_C=k_aN_AN_Bt_d/V$ is the microscopic change in the number of product molecules in step interval $t_d$ and $Z$ is the expected number of collisions between $A$ and $B$ in the interval (see Appendix \ref{alir}). The acceptance probability can then be expressed as \cite{Arjunan2010}
\eq{acpa}{P_a=\frac{k_a}{3\sqrt{2}(D_A+D_B)l}.}
The above relation is applicable when the reaction is activation-limited ($k_a\ll k_D$). For diffusion-influenced reactions ($k_a\gg k_D$), the collision rate $Z$ is reduced relative to the production rate $\Delta N_C$. The acceptance probability $P_a=\Delta N_C/Z$ would then have the issue of exceeding unity when $\Delta N_C > Z$. The Spatiocyte scheme overcomes this issue by reducing the simulation interval by a factor of $\alpha$ to $t'=t_d\alpha$. With the reduced interval, the effective number of collisions in $t_d$, $Z$ is increased. The step and reaction acceptance probabilities are then decreased accordingly to $P_w=\alpha$ and $P_a'=P_a\alpha$, respectively. Algorithm \ref{code} describes how $\alpha$ is set. In summary, the Spatiocyte scheme operates with $\alpha=1$ when $P_a\leq 1$ (activation-limited case) and with $\alpha <1$ when $P_a> 1$ (diffusion-influenced case).

\begin{algorithm}
\SetAlgoLined
  \caption{Basic outline of the Spatiocyte algorithm for bimolecular reactions. $t_{sim}$ is the current simulation time, $t_{end}$-$t_{sim}$ is the simulation duration, $P_{axy}$ is the reaction acceptance probability for the reactive pair of species $x$ and $y$, $t_{d}=l^2/6D_x$ is the diffusion (hopping) time step of the current species $x$, $l$ is the voxel size, $D_x$ is the diffusion coefficient of $x$, and $rand$ is a random number drawn from the uniform distribution with the interval $[0,1)$.}

\label{code}
Initialization:\\
$t_{sim} \leftarrow$0, scheduler $S \leftarrow \{\}$\\
\For{\normalfont{each species $x$}}{
$\rho_x=$max$\{P_{axy}\}$, where $xy$ denotes the pair of reactive species $x$ and $y$\;
$S\leftarrow t'=t_d\alpha$, where $\alpha=\left\{\begin{array}{lr}
1/\rho_x,& \text{for }\rho_x>1\\
1,& \text{for }\rho_x\leq1
\end{array}\right.$;\\
reaction acceptance probability $P'_{axy}=P_{axy}\alpha$\;
step acceptance probability $P_{wx}= \alpha$\;
}
Main loop:\\
\While{$S \neq \{\}$ \normalfont{and} $t_{sim}<t_{end}$}{
$t_{sim}\leftarrow \tau_x =$ next event in $S$;\\
get species identity $x$;\\
get current voxel location $s_{0}$;\\
reschedule next event, $\tau_x=\tau_x+t'$\\
\For{\normalfont{each molecule of species} $x$}{
choose a random target voxel $s_1\in$ $\{\text{nearest neighbor of }s_0\}$\;
	\uIf{$s_1$ \normalfont{is vacant}}{
		draw $rand$\;
		\lIf{$rand \leq P_{wx}$}{walk succeeded, $s_0\leftarrow s_1$}
         \Else{walk rejected, $s_0\leftarrow s_0$;}
         }
	\uElseIf{$s_1$ \normalfont{contains reactant species} $y$}{
	draw $rand$\;
		\uIf{$rand \leq P'_{axy}$}{reaction $xy$ accepted, $s_0\leftarrow s_1$}		 
		\Else{reaction failed and walk rejected, $s_0\leftarrow s_0$}
	}
	\Else{walk rejected, $s_0\leftarrow s_0$;}
}
}
\end{algorithm}

\subsection{Rebinding probability and reaction rate on HCP lattice}
As an alternative to the time-dependent reaction rate coefficient in Eq. \eqref{ktnoyes}, we define a discrete-space version with a step-dependent rate coefficient on lattice as  \cite{Torney1983,Montroll1969}
\eq{}{k_m = k_a'\left[1-\sum_{n=0}^m H_n\right],\ \text{for }m,n\in \mathbb{N},}
where $m$ is the simulation step, which is related to the simulation time by $6D_xt'=ml^2$, $k_a'$ is the initial reaction rate constant on lattice (see Appendix \ref{alir}) and $H_n$ is the lattice analogue of the rebinding-time probability function $p_{reb}(R,t|R,0)$ in diffusion step $n$. 
\\At long-time, the effective rate on lattice follows similarly to Eq. \eqref{longtimek}:
\eq{noyes}{k_{eff}'=\lim_{m\to\infty} k_m =k_a'\left[1-\sum_{n=0}^\infty H_n\right],}
where the summation term \eqref{noyes} corresponds to the total rebinding probability on lattice.

To obtain the analytical expression for $H_n$, we consider again a reactive pair $A$ and $B$, which are initially in-contact by occupying adjacent voxels on lattice. We are interested in the rebinding-time probability distribution as a function of the diffusion step $n$. Without losing generality, we can fix one of the molecules and diffuse the other with the relative diffusion coefficient $D$. Then, the rebinding-time probability distribution of $A$ and $B$ is related to the arrival-time probability distribution of a random walker to the origin for the first time, given that the walk started at one of the neighbor voxels of the origin with diffusion coefficient $D$. In the following sections, we define $H_n$ explicitly and use it to derive the rate coefficient on HCP lattice. Since the approaches for activation-limited and diffusion-influenced cases are different in the Spatiocyte scheme, we perform their derivations separately.

\subsubsection{Activation-limited case ($k_a\ll k_D$, $\alpha=1$)}
We denote $s_0$ as the voxel at origin, $s_1$ as an element of the set of immediate neighbor voxels of $s_0$. We define $F_n(s_a|s_b)$ as the first-passage time distribution for a random walker to walk from voxel $s_b$ to $s_a$, that is, the probability of arriving at voxel $s_a$ for the first time at the $n$th step, given that the walk started at voxel $s_b$.  

We first consider the rebinding-time probability distribution for the case $P_a=1$.  Let $F_n(s_0 | s_0)$ and $F_n(s_0|s_1)$ denote the first-passage time distributions to origin from origin and $s_1$, respectively. The two probabilities are related via 
\eq{}{p(s_0\rightarrow s_1)F_n(s_0 | s_1)=F_{n+1}(s_0 | s_0),}
where $p(s_0\rightarrow s_1)=1$ is the transition probability from $s_0$ to $s_1$ in a single step. This implies that the trajectory we are interested in, which is from an in-contact situation (e.g., $A$ at $s_1$ and $B$ at $s_0$) to the rebinding situation ($A$ hops to $s_0$) in a single step, is equivalent to the 2-step trajectory, $s_0\to s_1\to s_0$. \\

Therefore, the rebinding-time probability distribution is fully described by $F_n(s_0|s_1)$ and is related to $F_n(s_0|s_0)$. The latter can be obtained analytically from its probability generating function $F(s_0|s_0;z)=\sum_{n=0}^\infty F_n(s_0|s_0)z^n$ \cite{joyce98} (see Appendix \ref{aftd}).

As for $P_a\leq1$, the trajectories that have undergone failed reaction attempts before step $n$ are included in the rebinding-time probability distribution:
\eq{HN}{H_n(s_0|s_1) = P_a \sum_{j=1}^{n} F_{n+j}^{j}(s_0|s_0)(1-P_a)^{j-1},\text{ for }n\in \mathbb{N},\ j \in \mathbb{Z}^+,}
where $F^j_n(s_0|s_0)$ is the probability to reach the origin for the $j$th time at the $n$th step (I.1.9 in \cite{montroll1965random}):
\eq{}{F_n^j(s_0|s_0)=\sum_{i=1}^{n}F_{n-i}^{j-1}(s_0|s_0)F_i(s_0|s_0),\text{ for }\ j \in \mathbb{Z}^+,}
where $F_n^1(s_0|s_0)=F_n(s_0|s_0)$.

The generating function of $H_n(s_0|s_1)$ in terms of $F(s_0|s_0;z)$ is (see Appendix \ref{aalc}):
\eq{gfal}{H(s_0|s_1;z) = \frac{P_aF(s_0|s_0;z)}{F(s_0|s_0;z)(P_a-1)+z}.}
\\By taking the limit $z\to 1$ on $H(s_0|s_1;z)$, we obtain the total rebinding probability on lattice as
\eq{Htot}{
H_{reb}=\lim_{z\to 1}H(s_0|s_1;z)=\frac{P_a}{P_a+\frac{1}{F(1)}-1},}
where $F(1)=F(s_0|s_0;z=1)$. It was shown previously that the probability generating function of the HCP lattice is topologically equivalent to that of the face-centered cubic (FCC) lattice \cite{Ishioka1978}. Therefore, we have $F(1)\approx 0.256318$ (\cite{hughes1995}, p. 153) for HCP lattice. 

Finally, if we set the initial rate $k_a' =3\sqrt{2}lDP_a$ (see Appendix \ref{alir}) and substitute the total rebinding probability $H_{reb}$ from Eq. \eqref{Htot} into Eq. \eqref{noyes}, we obtain the effective rate constant on lattice as
\meq{Eff}{
k_{eff}'&=3\sqrt{2}Dl \left(\frac{1}{F(1)}-1\right) \frac{P_a}{P_a+\frac{1}{F(1)}-1}\ .
}

\subsubsection{Diffusion-influenced case ($k_a\gg k_D$, $\alpha<1$)}\label{dlim}
The rebinding-time probability distribution $G_n(s_0|s_1)$ of the diffusion-influenced scheme is defined as
\eq{}{G_{n+1}(s_0|s_1)=S_n(s_1|s_1)\ p(s_1\to s_0),\text{ for }n\in \mathbb{N},}
where 
\eq{}{p(s_1\to s_0)=\frac{P_a\alpha P_1(s_0|s_1)}{1-(1-P_w)(1-P_1(s_0|s_1))}}
is the probability for a successful reaction, $P_1(s_0|s_1)$ is the probability to select $s_1$ given that the molecule is in $s_0$ ($=\frac{1}{12}$ for HCP lattice), and $S_n(s_1|s_1)$ is the probability that a particle is in contact after $n$-steps (see Appendix \ref{adic}  for more details).
\\The probability generating function of $G_n(s_0|s_1)$ on HCP lattice is given by (see Appendix \ref{adic})
\eq{gfdi}{G(s_0|s_1;z)=\frac{P_a\alpha /12}{1-11(1-\alpha)/12}S(s_1|s_1;z),}
where $S(s_1|s_1;z)$ is the probability generating function of $S_n(s_1|s_1)$ .

Taking the limit $z\to 1$, we get the total rebinding probability as (Appendix \ref{adic} )
\eq{Greb}{G_{reb}=\lim_{z\to 1}G(s_0|s_1;z)=\frac{P_a}{P_a+\frac{1}{F(1)}-1},}
which is identical to Eq. \eqref{Htot} in the activation-limited case.
Similarly, by substituting the summation term in Eq. \eqref{noyes} with Eq. \eqref{Greb}, we get the effective rate constant for the diffusion-influenced case as
\meq{latkef}{
k_{eff}'&=k_a'[1-G_{reb}],\\
}
which also follows Eq. \eqref{Eff}. Henceforth, we adopt the same notations of the effective reaction rate and total rebinding probability for both the activation-limited and diffusion-influenced cases.

\subsection{Comparison with continuum-based theory}
Since the effective rate on lattice \eqref{latkef} has the same form of Eq. \eqref{keff2} in continuum, we can match them by equating the initial rate and total rebinding probability of the two: $k_a'=k_a$ and $G_{reb}=P_{reb}$. With the former relation, the reaction acceptance probability is connected to the initial rate constant, diffusion coefficient, and voxel size by
\eq{pa}{ P_a= \frac{k_a}{3\sqrt{2}Dl}.} 
Employing the $G_{reb}=P_{reb}$ relation, the voxel size is found to be about $2\%$ greater than the molecule size:
\eq{lfactor}{
l=\frac{4\pi R}{ 3\sqrt{2}(\frac{1}{F(1)}-1)}\approx 1.0209R.
}
The Spatiocyte scheme is thus guaranteed to have the same effective rate and total rebinding probability as the continuum framework provided that Eqs. \eqref{pa} and \eqref{lfactor} are satisfied. In addition, the expression for lattice effective rate constant follows the same form of the continuum-based framework:
\meq{}{
k_{eff}'&=\frac{k_a'k_D'}{k_a'+k_D'}=k_D'G_{reb},\\
G_{reb}&=\frac{1}{1+k_D'/k_a'},
} 
where $k_D'=3\sqrt{2}lD ({1}/{F(1)}-1)$.

According to Eq. \eqref{lfactor}, accurate matching of both the effective rate and the total rebinding probability requires the voxel size to be larger than the molecule size. Nonetheless, during modeling we can fix the voxel size to be the same as the molecule size, $l=R$. In this case, it is still possible to match the lattice effective reaction rate to the continuum-based rate by setting the reaction acceptance probability to
\eq{newpa}{
P_a=(1/F(1)-1)\left[\frac{3\sqrt{2}(k_a+k_D)(1/F(1)-1)}{4\pi k_a}-1\right]^{-1}.
}
However, this is done at the expense of losing accuracy in the total rebinding probability, since $G_{reb}\neq P_{reb}$.

For the reversible reaction, $A+B \underset{k_d}{\stackrel{k_a}{\rightleftharpoons}} C$, the local detailed balance on lattice is achieved by choosing a lattice dissociation rate constant $k_d'$ from the following equilibrium constant relation:
\eq{}{K_{eq}=\frac{k_a'}{k_d'}=\frac{k_a}{k_d}.}
The MLM method can simulate the dissociation reaction as a first-order process with rate $k_d'$ and place the dissociated pair of molecules at an in-contact condition.

\subsection{Numerical simulations}
We verify the main theoretical results presented above with numerical simulations using Spatiocyte. Spatiocyte is included in E-Cell System version 4 \cite{ecell4}, an open-source biochemical simulation environment that supports multiple algorithms, time scales and spatial representations. The Python notebooks used to generate the simulation results reported here are available at \url{https://github.com/wxchew/MLM}. The performance benchmark models for all tested methods are included in Spatiocyte package (\url{http://spatiocyte.org}) as examples.

\section{Results and Discussion}
We first validate the theory of total rebinding probability and its time-dependent behavior on lattice using numerical simulations. We examine the accuracy of the reaction rate coefficient and its time-dependent behavior on lattice. We then compare the diffusion and reaction performances of MLM and several other off-lattice particle methods. Finally, we evaluate MLM in protein production-degradation process, dual phosphorylation cycle and a simple bimolecular reaction in a crowded compartment.

\subsection{Numerical validation of MLM theory}
\subsubsection{Rebinding probability}
We examine the rebinding probability distribution of a reactive pair, $A$ and $B$ that are initially in-contact. The theoretical rebinding-time probability distribution  $H_n(s_0|s_1)$ and $G_n(s_0|s_1)$ are validated against numerical results. In the activation-limited case ($k_a/k_D\ll1$), the expected first rebinding probability at $n$th step is obtained using Eq. \eqref{HN}, whereas in the diffusion-influenced case ($k_a/k_D\geq1$), the probability is calculated from the generating function $G(s_0|s_1;z)$
\eq{}{
G_n(s_0|s_1)= \left. \left[\frac{1}{n!}\frac{d^n}{dz^n}G(s_0|s_1;z)\right]\right|_{z=0}.
}
Table \eqref{hngn} shows the simulated and the expected theoretical values for $n\in [1,5]$ steps. The simulation results agree well with the expected values, with discrepancies never exceeding $0.1\%$. Since the theoretical rebinding-time probability distribution on lattice is validated by simulations, the analytical formulas for the total rebinding probability derived from it, Eqs. \eqref{Htot} and \eqref{Greb} are therefore valid.

To illustrate the dependency of total rebinding probability on $k_a/k_D$, we obtained the probability at various $k_a/k_D$ up to $n=10$. Table \eqref{t1} shows the simulated and the expected theoretical values for various $k_a/k_D$ ratios. Both simulated and theoretical values coincide well, with discrepancies never exceeding $0.03\%$. Qualitatively, the total rebinding probability increases with larger $k_a/k_D$ values, consistent with the continuum theory \eqref{ptot}. 

\begin{table}[]
	\caption{Theoretical and simulated rebinding-time probabilities on lattice for activation-limited and diffusion-influenced cases. Simulation parameters: $l=0.01\ \mu \mathrm{m}$, volume = $(100\ l)^3$ with periodic boundary, runs = $1\times10^9$.}
\label{hngn}
\centering
\begin{tabular}{ l l l l l l l }
\hline
\hline
&\multicolumn{3}{c}{$H_n(s_0|s_1)$, $P_a=0.5$} &\multicolumn{3}{c}{$G_n(s_0|s_1)$, $P'_a=2\alpha$, $\alpha=1/2$}\\
n		&Theory  &Simulation  &Error (\%) &Theory  &Simulation  &Error (\%)\\
\hline
1		&0.0416666 	&0.0416586 &0.019 &0.1538461 &0.1538326 &0.009\\
2		&0.0156250 	&0.0156228 &0.014 &0.0473373 &0.0473431 &0.012\\
3		&0.0107784	&0.0107779 &0.005 &0.0313306 &0.0313317 &0.004\\
4		&0.0074297	&0.0074274 &0.031 &0.0200584 &0.0200534 &0.026\\
5		&0.0056802	&0.0056773 &0.049 &0.0147588 &0.0147496 &0.062\\
\hline
\hline
\end{tabular}
\end{table}

\begin{table}[]
\caption{Theoretical and simulated rebinding probabilities up to $n=10$ steps with $k_a/k_D$ ratios ranging from the highly activation-limited case ($k_a/k_D=0.01$) to the strongly diffusion-influenced case ($k_a/k_D=100$). Simulation parameters: $l=0.01\ \mu \mathrm{m}$, $D=1\ \mu \mathrm{m^2/s}$, volume = $(10000\ l)^3$ with periodic boundary, runs = $1\times10^9$. $\alpha=1/P_a$ for diffusion-influenced cases.}
\label{t1}
\centering
\begin{tabular}{ l *{5}{l} }
\hline
\hline
$k_a/k_D$	&0.01 & 0.1 & 1 & 10 & 100\\
\hline
Lattice theory				&0.0062657 &0.05973879 &0.397486 &0.874154 &0.985988		\\
Simulation 				 	&0.0062672 &0.05973410 &0.397459 &0.874126 &0.986000			\\
Discrepancy ($\%$) 			&0.025     &0.0078 	   &0.0068   &0.0032   &0.0012			\\
\hline
\hline
\end{tabular}
\end{table}

\begin{figure}[!h]   
\centering
  \includegraphics[width=0.5\textwidth]{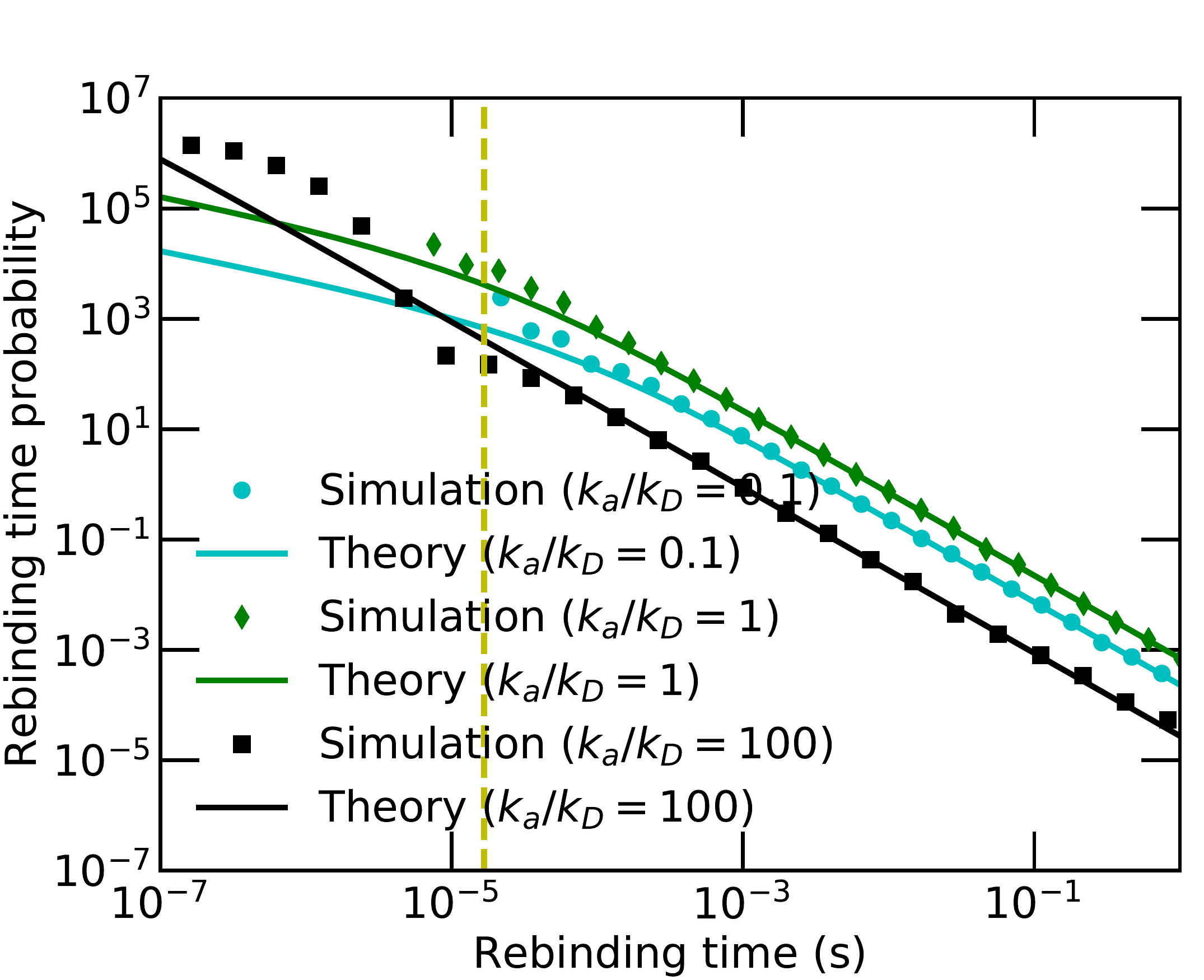}  
  \vspace{0.4cm}
  \caption{ The rebinding time of a reactive pair that is initially in-contact. The rebinding time is sampled from simulations with $k_a/k_D$ = 0.1, 1 and 100. Markers show the simulation results of Spatiocyte while solid lines depict the analytical results from the continuum-based scheme \eqref{prebt}. The vertical dashed line marks the characteristic diffusion time step, $t_d$. Simulation parameters: $l = 0.01\ \mu \mathrm{m}$, volume = $(10000\ l)^3$ with periodic boundary, runs = $10^4$, $D_A = 1\mu \mathrm{m^{2} s^{-1}}$, $D_B=0$, $\alpha=1/P_a$ for the diffusion-influenced case.
  }
\label{rebinding}
\end{figure}

We then evaluated the rebinding-time probability distribution by recording the time taken for $A$ and $B$ to associate immediately after a dissociation event. We performed the simulations for a large number of steps and independent runs. Figure \ref{rebinding} shows the average number of rebinding events per unit time at $k_a/k_D=0.1, 1$ and $100$. Lines depicting the rebinding-time probability distribution of the continuum-based model according to Eq. \eqref{prebt} are also shown as reference. It is clear that at times larger than $t_d$, the time-dependent behavior of lattice simulations is consistent with the continuum-based model. The scaling behavior at long-time, $p_{reb}(t) \propto t^{-3/2}$ is a well-known characteristic of a 3D random walker returning to the origin \cite{pfluegl1998}. We have corroborated this result with detailed asymptotic analysis that is provided in Appendix \ref{arplt}.

Note that in the diffusion-influenced case ($k_a/k_D=1\text{ and }100$), finer step intervals generate rebindings at times smaller than the diffusion time step $t_d$, denoted by the vertical dashed line in Figure \ref{rebinding}. In this temporal regime, MLM behaves differently from the continuum-based framework because the MLM reaction kinetics approximates the Poisson process (see Appendix \ref{atdsp}). Despite the difference, the rebinding behavior correctly converges to the continuum-based formalism for times larger than $t_d$.

\subsubsection{Reaction rate}

\begin{figure}[!h] 
\centering
    \begin{subfigure}[t]{0.49\textwidth}
  	\centering
    \includegraphics[width=\textwidth]{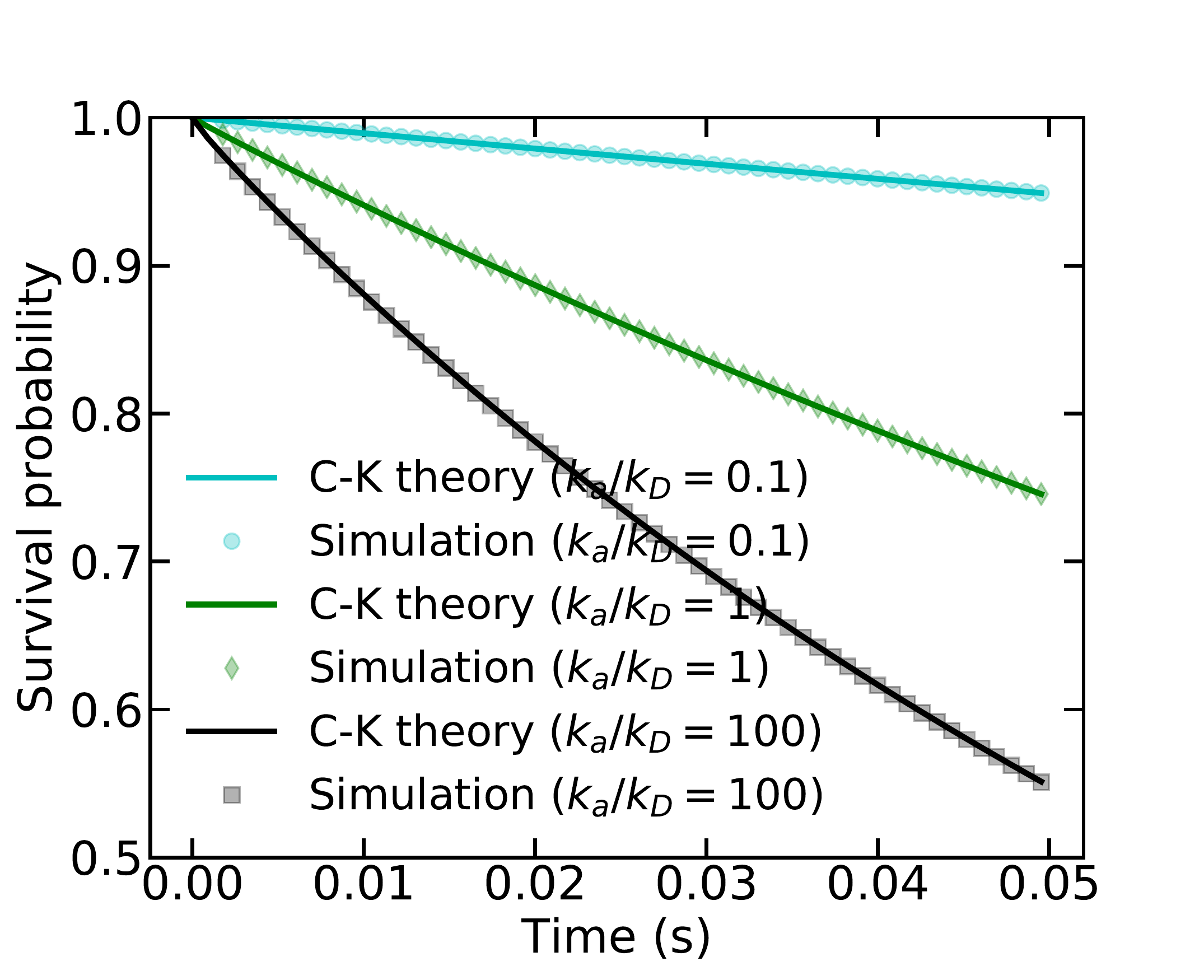}    
  	\caption{}
	\label{fig:survival}
	\end{subfigure}
	\hfill
	\begin{subfigure}[t]{0.49\textwidth}
   	\centering	  
    \includegraphics[width=\textwidth]{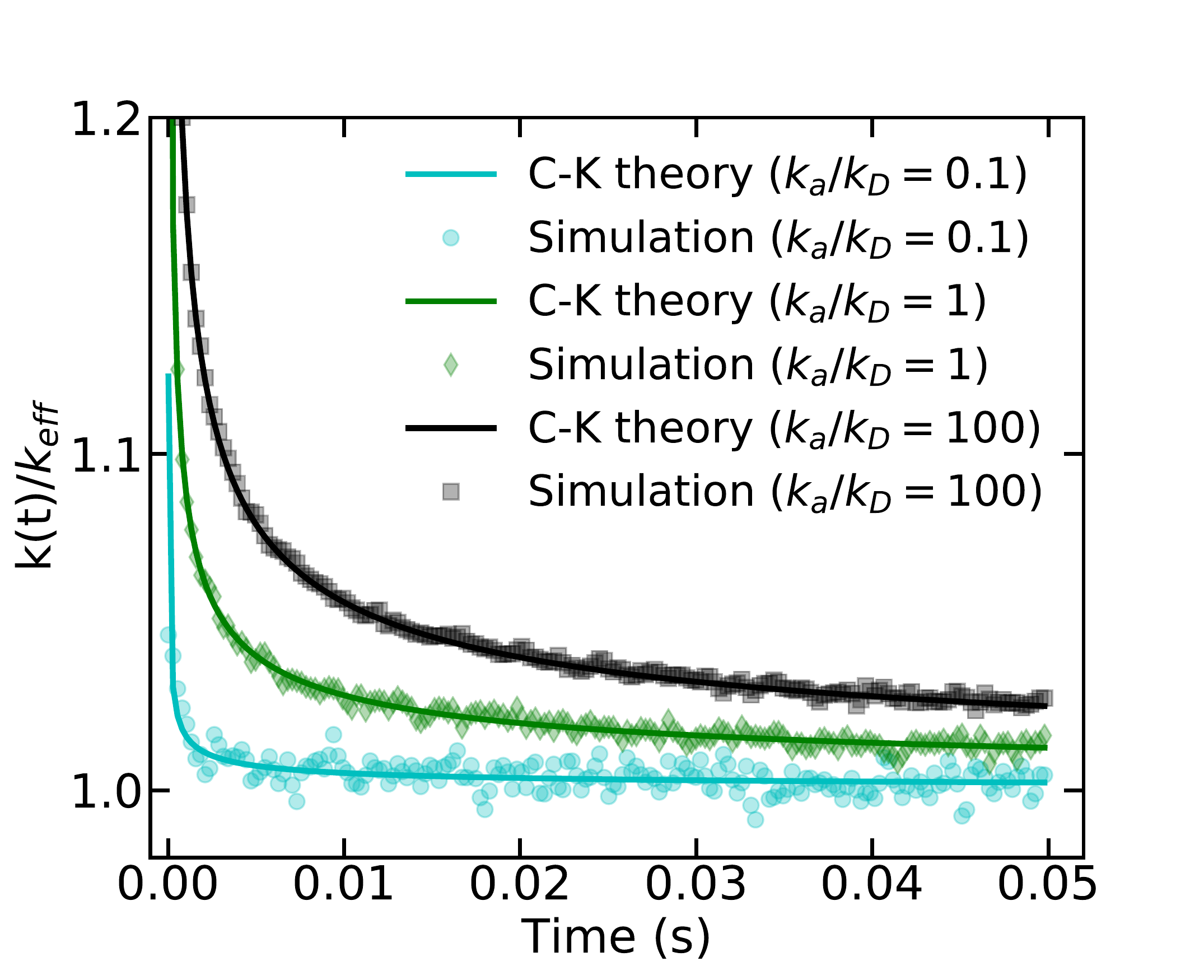}    
  	\caption{}
	\label{fig:kt1}   	
	\end{subfigure}
	
	\caption{Survival probability and time-dependent rate coefficient. (a) Survival probability of $A$ in $A+B \xrightarrow{} B$ with $k_a/k_D$ = 0.1, 1 and 100. (b) Simulated time-dependent rate coefficients of the reaction and the corresponding long-time approximation of Collins-Kimball (C-K) theory in Eq. \eqref{klarget}. Simulation parameters: volume = $(3.5\ \mu \mathrm{m})^3$ with periodic boundary, $R=0.01\ \mu \mathrm{m}$, $l=0.01\times1.0209\ \mu \mathrm{m}$, $D_A=0$, $D_B=1\ \mu \mathrm{m^{2} s^{-1}}$, $N_a=N_b=4000$, duration = 0.05 s, runs = $3\times 10^4$, $\alpha=1/P_a$ for the diffusion-influenced case.}
\end{figure}

We evaluated the accuracy of the effective reaction rate constant for irreversible bimolecular reactions \eqref{birxn} over various $k_a/k_D$ regimes on lattice. We considered an immobile species $A$ and a diffusing species $B$ that are uniformly distributed at initialization with concentrations $[A]$ and $[B]$, respectively. We recorded the surviving fraction of $A$ molecules at each time step. Figure \ref{fig:survival} displays the survival probability of $A$ and the expected theoretical curve $S_A(t)=\exp[-[B]\int_0^tk_{irr}(t')dt']$ (Eq. 2.35 in \cite{szabo1989}). From the survival probability, we calculated the time-dependent reaction rate coefficient using (Eq. 2.1 in \cite{szabo1989})
\eq{}{k_{irr}(t)=-\frac{1}{[B] S_A(t)} \frac{dS_A(t)}{dt}.}
We adopted the following discretization scheme for the time derivative to get the discrete rate coefficient:
\eq{ktcal}{k_{n+1}=-\frac{S_{n+2}-S_{n}}{[B]S_{n+1}\ (t_{n+2}-t_{n})},\text{ for }n\in \mathbb{Z}^+,} 
where $n$ is the index of the discretized $S_A$ and $t$. The boundary cases are computed as
\eq{}{k_1=-\frac{S_{2}-S_{1}}{[B]S_{1}\ (t_2-t_1)},\ k_N=-\frac{S_{N}-S_{N-1}}{[B]S_{N}\ (t_{N}-t_{N-1})},}
where $N$ denotes the final time step.
The reaction rate coefficient obtained for various $k_a/k_D$ ratios are shown in Figure \ref{fig:kt1} along with their corresponding theoretical curves from Eq. \eqref{klarget}. 

Recall that the long-time asymptotic variant of the Collins-Kimball theory \eqref{klarget} has the form
\eq{kfit}{k_{irr}(t)\simeq C_1\left(1+\frac{C_2}{\sqrt{t}}\right),}
where $C_1$ and $C_2$ denote the steady-state rate constant and the time-dependent term, respectively. We fitted Eq. \eqref{kfit} to the numerical data, omitting early time points to avoid non-steady-state effects. The resulting $C_1$ and $C_2$ parameters after fitting are listed in Table \ref{t2}. The theoretical values correspond favorably to the estimated steady-state reaction rate constants and are well within the standard error, thus validating the lattice theory for the effective rate. The time-dependent terms are also in good agreement with the theory, especially in the diffusion-limited case, with discrepancy less than $1\%$. This is consistent with the asymptotic analysis carried out in Appendix \ref{arclt} and \ref{arcalt}. In the activation-limited case ($k_a/k_D=0.1$), the fitted $C_2$ had the largest deviation from theory because the standard error was also the highest. The low number of data points contributed to the higher standard error. Nonetheless, we did not increase the data points because $C_2$ has a weaker influence in activation-limited reactions than $C_1$.

\begin{table}[h]
	\caption{The steady-state rate constant $C_1$ and the time-dependent term $C_2$ of reaction \eqref{birxn} at various $k_a/k_D$ were obtained by fitting the simulated reaction rate coefficient with Eq. \eqref{kfit}. Uncertainty in the simulated data was used as a weight in the fitting. Theoretical values from Eq. \eqref{klarget} are listed for comparison. Simulation parameters: $l=0.01\times1.0209\ \mu \mathrm{m}$, $D_A=0$, $D_B=1\ \mu \mathrm{m^2s^{-1}}$, volume = $(350\ l)^3$ with periodic boundary, $N_A=N_B=4000$, duration = $0.05\ \mathrm{s}$, runs = $3\times 10^4$, $\alpha=1/P_a$ for the diffusion-influenced case.}
\label{t2}
\centering
\begin{tabular}{ l *{4}{l} }
\hline
\hline
$k_a/k_D$							& 0.1  						& 1 		 		&100\\
\hline
Theoretical $C_1$ ($\mu \mathrm{m^3s^{-1}}$)	&0.011424				&0.062832				&0.124420	\\
Simulation 			       			&0.011423$\pm$0.0012 	&0.062848$\pm$0.0029 	&0.124459$\pm$0.0046	\\
Discrepancy ($\%$)					&0.011					&0.026					&0.032		\\
\hline
\hline
Theoretical $C_2$ ($\mathrm{s^{1/2}}$)	&0.00051 				&0.00282 				&0.00559	\\
Simulation 						&0.00054$\pm$0.01 		&0.00279$\pm$0.0052 	&0.00563$\pm$0.004	\\
Discrepancy ($\%$)				&5.5   					&1.04   				&0.77			\\
\hline
m.s.e. of fit					&3.4$\times 10^{-7}$ 	&2.2$\times 10^{-6}$ 	&4.2$\times 10^{-6}$	\\
\hline
\hline
\end{tabular}
\end{table}

\subsection{Performance}
\subsubsection{Diffusion}
\begin{figure}[!h]   
\centering
    \begin{subfigure}[t]{0.49\textwidth}
  	\centering
    \includegraphics[width=\textwidth]{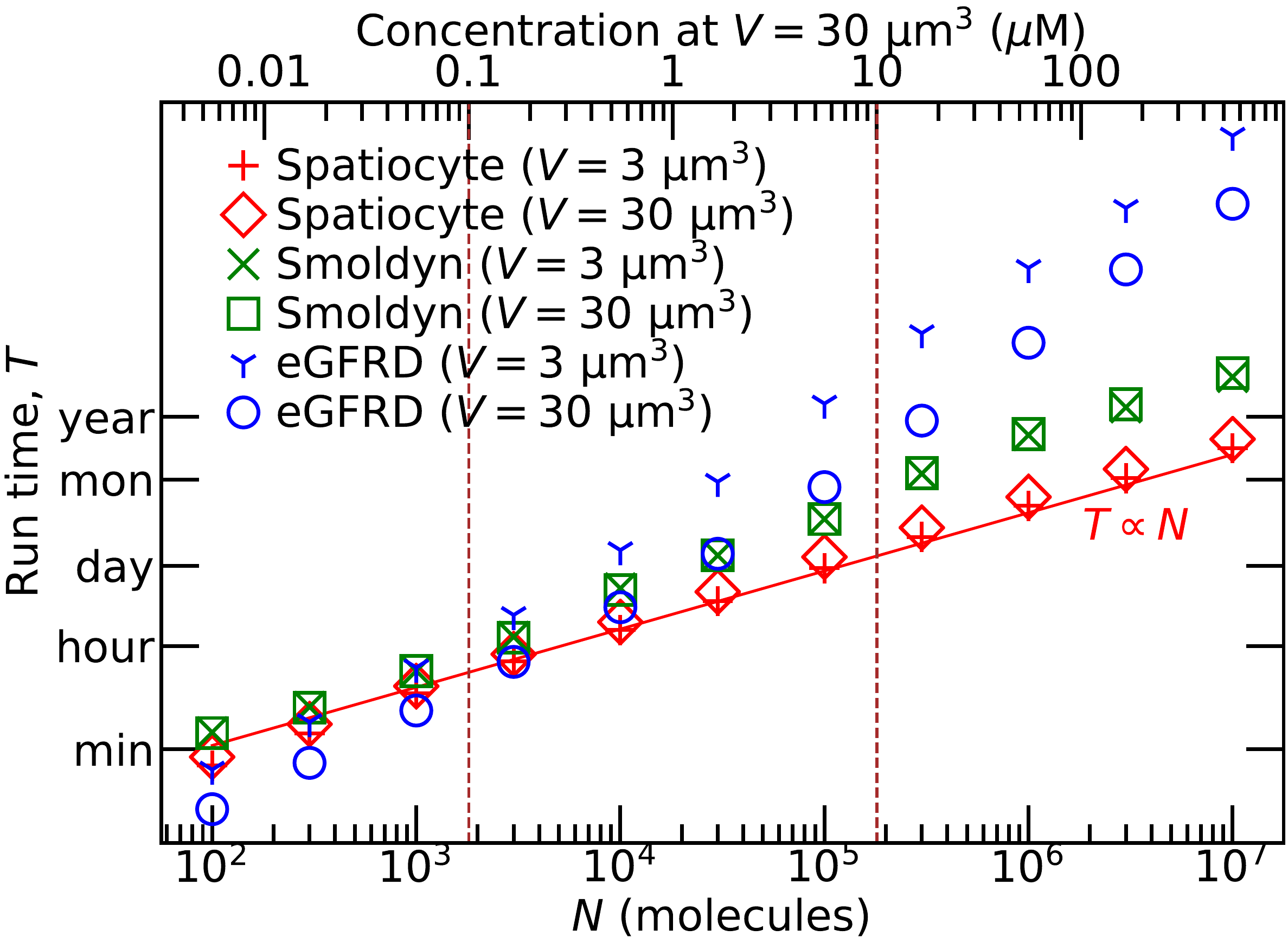}    
  	\caption{}
	\label{fig:perfsphere}
	\end{subfigure}
	\hfill
	\begin{subfigure}[t]{0.49\textwidth}
   	\centering	  
    \includegraphics[width=\textwidth]{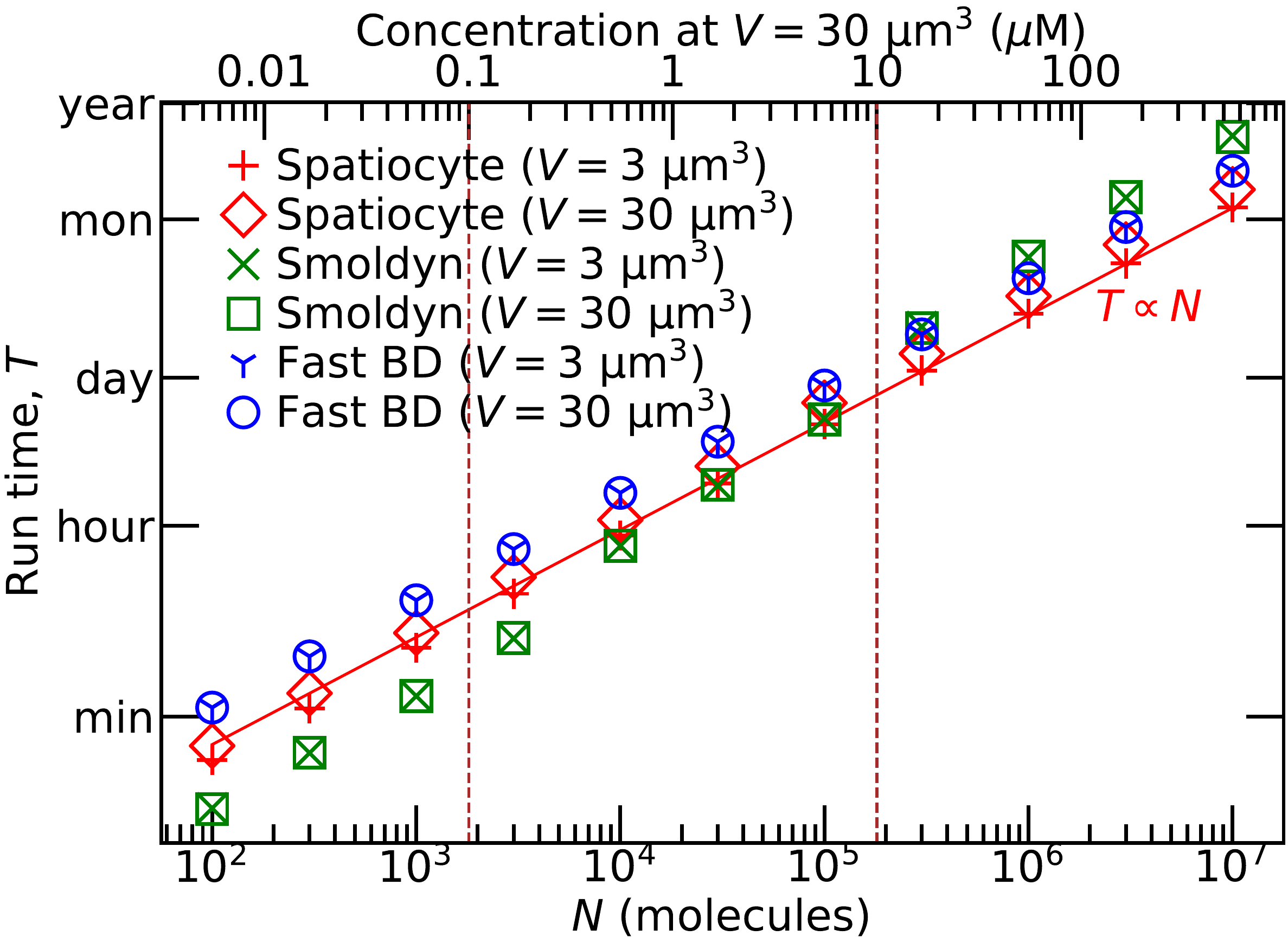}    
  	\caption{}
	\label{fig:perfpoint}   	
	\end{subfigure}

  \caption{3D diffusion performance of particle-based methods. Vertical axis, $T$ shows the run times to diffuse molecules with diffusion coefficient $D_x$ = 1 $\mathrm{\mu m}^2\mathrm{s}^{-1}$ in volume $V$ for 10 s. Bottom horizontal axis, $N$ represents the number of diffusing molecules, while the top axis shows the corresponding concentration at $V=30\ \mathrm{\mu m}^3$. (a) Molecules are represented as hard-sphere particles with radius $r=l/2$ = 2.5 nm. (b) Molecules are dimensionless point particles that can overlap one another. eGFRD does not support point particle diffusion and conversely, fast BD here can only diffuse point particles. Smoldyn simulation interval is set to the step interval $t_d$ (4.17 $\mu$s) of Spatiocyte and fast BD for comparison. The eGFRD algorithm uses variable time steps. Each model was simulated for a predefined run time, $t_r$ and the resulting simulated time, $t_s$ was recorded. We calculated $T$, the run time in seconds for 10 s of simulated time with $T=10t_r/t_s$. $t_r$ was set such that at least hundreds of simulation steps have been completed. The resulting range of $t_r$ was between 1 hour to several days. Solid lines depict the ideal scaling for Spatiocyte. Vertical dashed lines indicate the typical concentration range of proteins in the cytoplasm (0.1 to 10 $\mu$M). All simulations were executed on the same server with Intel Xeon Platinum 8180 2.5 GHz (max 3.80 GHz) CPU, 768 GB memory and Ubuntu 18.04 LTS operating system.
  }
\label{fig:performance}
\end{figure}

We compared the 3D diffusion performance of MLM using Spatiocyte (git 9757fb3) and three other off-lattice particle-based simulation methods, Smoldyn \cite{andrews2010} (version 2.55), eGFRD \cite{takahashi2010} (in E-Cell System version 4.1.4) and fast Brownian dynamics (BD) \cite{Smith2017} (C++ program example in Spatiocyte git 9757fb3). When the molecules are represented as hard-spheres with volume exclusion, Spatiocyte required shorter run times than Smoldyn in all cases (Figure \ref{fig:perfsphere}). Spatiocyte achieves comparable or better performance than eGFRD in the typical concentration range of cytoplasmic macromolecules (0.1 to 10 $\mu$M). For example at 6 $\mu$M in volume $30\ \mathrm{\mu m}^3$, Spatiocyte is about 4.5 and 16 times faster than Smoldyn and eGFRD, respectively. In contrast to eGFRD, Spatiocyte and Smoldyn execution times increase with the number of molecules but not the molecular crowdedness ($V=30\ \mathrm{\mu m}^3$ vs. $3\ \mathrm{\mu m}^3$). The simulation times of Spatiocyte scale almost linearly with the number of molecules ($T\propto N$), which is not apparent with Smoldyn and eGFRD. The drastic slowdown of eGFRD at higher concentrations is caused by the shorter time steps required to resolve many molecular interactions that take place in the densely occupied system \cite{takahashi2010}. 

If molecules are represented as dimensionless point particles, higher diffusion performance is expected since inter-molecular collisions can be ignored. Figure \ref{fig:perfpoint} shows the run times of Spatiocyte, Smoldyn and fast BD when diffusing point particles with the same simulation interval. eGFRD was not considered here since it only supports molecules with physical volume. Spatiocyte and fast BD execution times showed an almost linear scaling with the number of molecules. Although Smoldyn did not scale as well, it had the fastest run times when the number of diffusing molecules was 30,000 or less. Spatiocyte outperformed fast BD in all tests and is on average 2.5 times faster. As expected, the simulation times of all three methods were not affected by the crowdedness in the volume since molecular collisions are disregarded. On average, Spatiocyte takes about 2 times longer to diffuse hard-sphere molecules than point particles.

\subsubsection{Reaction}
\begin{figure}[!h]   
\centering
  \includegraphics[width=0.5\textwidth]{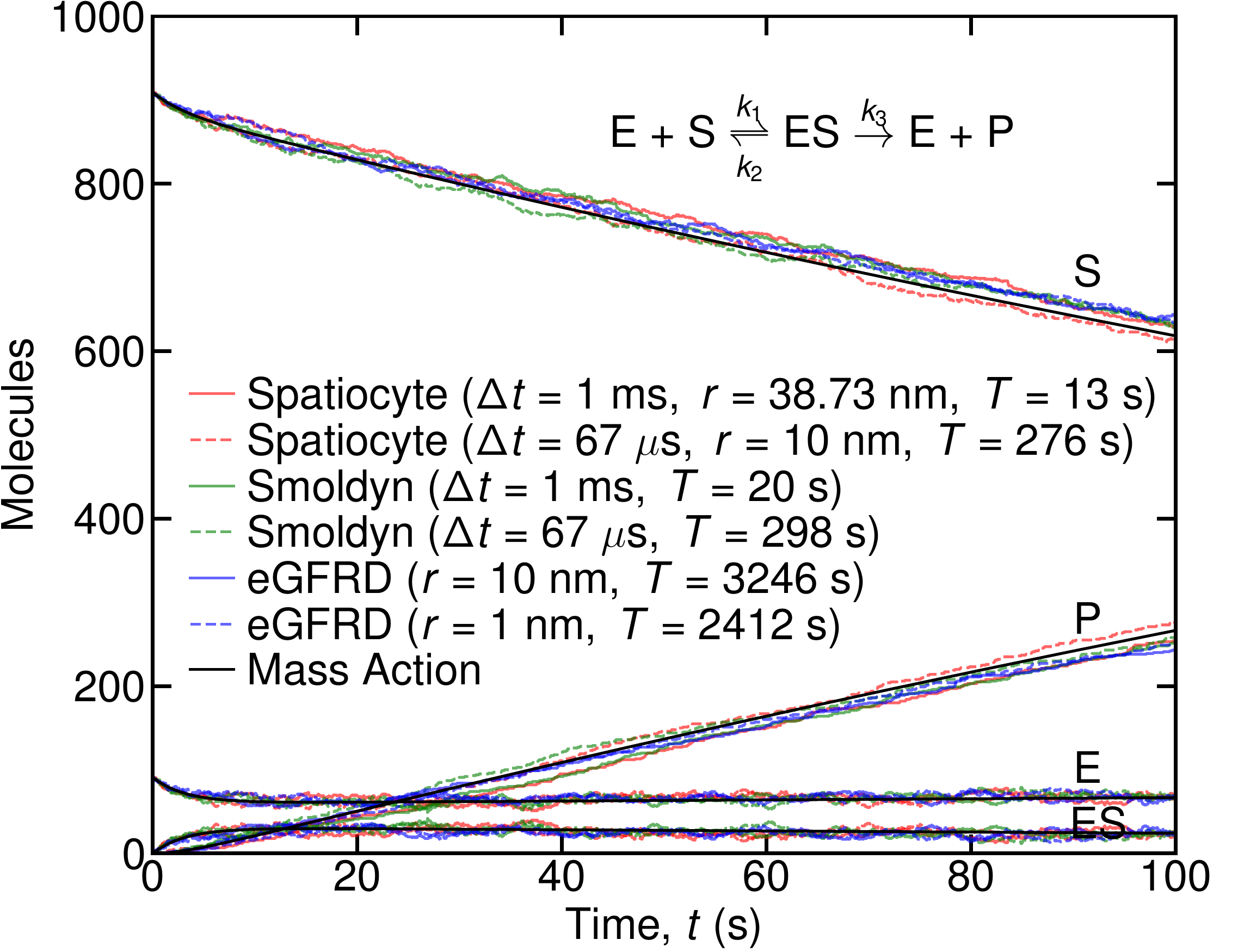}  
  \vspace{0.4cm}
  \caption{Particle simulation performance of the Michaelis-Menten reaction. Original benchmark model from \cite{andrews2010, Andrews2018} was modified with volume ($V$) = 90.9 $\mu \mathrm{m}^3$, diffusion coefficient ($D_x$) = 1 $\mu\mathrm{m}^2\mathrm{s}^{-1}$, $k_1$ = 0.01 $\mu \mathrm{m}^3\mathrm{s}^{-1}$, $k_2$ = $k_3$ = 0.1 $\mathrm{s}^{-1}$. Molecule or voxel radius ($r$), simulation or diffusion step interval ($\Delta t$) and run time ($T$) are as indicated. All simulations were executed on the same workstation with Intel Xeon X5680 3.33 GHz CPU, 48 GB memory and Ubuntu 16.04 LTS operating system.
  }
\label{fig:perfreact}
\end{figure}

Recently, Andrews \cite{Andrews2018} benchmarked the performance of Smoldyn, MCell \cite{Kerr2008}, eGFRD, SpringSaLaD \cite{michalski2016} and ReaDDy \cite{schoneberg2013} particle simulators when running the well-known Michaelis-Menten enzymatic reaction. Smoldyn required the least amount of time to complete the benchmark. Running the model on our hardware (see Figure \ref{fig:perfreact} for specifications) with the same 1 ms simulation interval, Spatiocyte took 113 s, whereas Smoldyn required 31 s. Since it would take too long for eGFRD to complete the simulation of the original model \cite{Andrews2018}, we decreased the number of molecules, diffusion coefficients and reaction rates. The execution times of Spatiocyte, Smoldyn and eGFRD when running the model with the new parameters are presented in Figure \ref{fig:perfreact}. The simulators generated almost identical results. Spatiocyte and Smoldyn had similar run times, whereas eGFRD required about one to two orders of magnitude longer. Although Spatiocyte is about four times slower than Smoldyn when executing the original model, both had very similar times with the new parameters. Our results indicate that the relative performance of Spatiocyte and Smoldyn depends on the model parameters. 

\subsection{Application Examples}
We applied MLM to model two fundamental RD systems of intracellular signaling, the production-degradation process, previously studied using lattice-based methods \cite{erban09,Sturrock2016, cianci2016}, and the dual phosphorylation-dephosphorylation cycle of the mitogen-activated protein kinase (MAPK) cascade \cite{chang2001mammalian,ferrell1997,aoki2011}, a common motif found in signal transduction systems but with a response function that is highly sensitive to the binding kinetics. We also report the effects of excluded volume on the kinetics of a simple bimolecular reaction using MLM.

\subsubsection{Production-degradation process}
\begin{figure}[!h] 
\centering
	\begin{subfigure}[t]{0.49\textwidth}
   		\centering	  
	    \includegraphics[width=\linewidth,height=7cm]{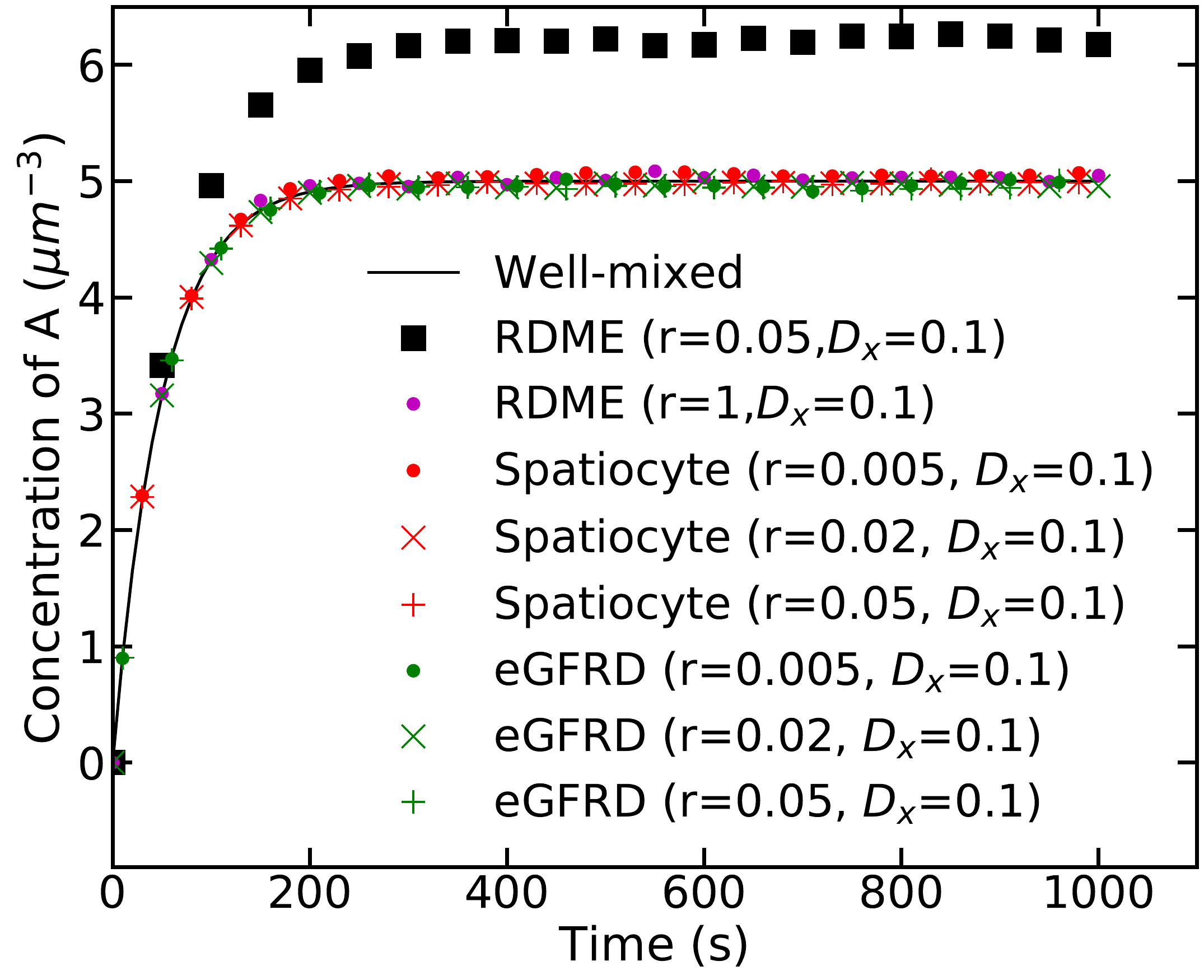}    
  		\caption{}
		\label{fig:Acurve}
	\end{subfigure}
	\hfill
    \begin{subfigure}[t]{0.49\textwidth}
  		\centering
	    \includegraphics[width=\linewidth,,height=7cm]{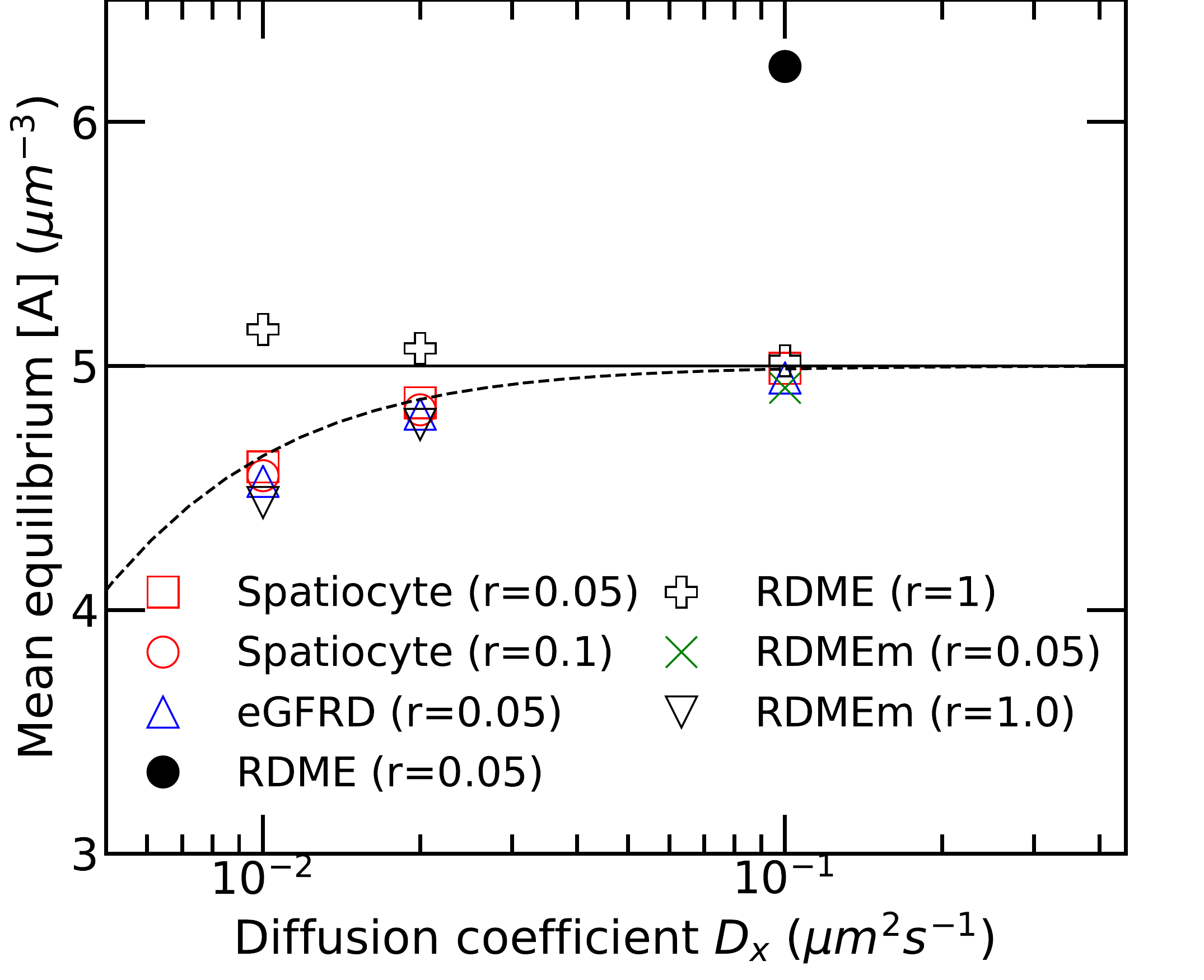}    
  		\caption{}
		\label{fig:Ddpnt}
	\end{subfigure}
    \begin{subfigure}[t]{0.6\textwidth}
  		\centering
	    \includegraphics[width=9cm,height=7cm]{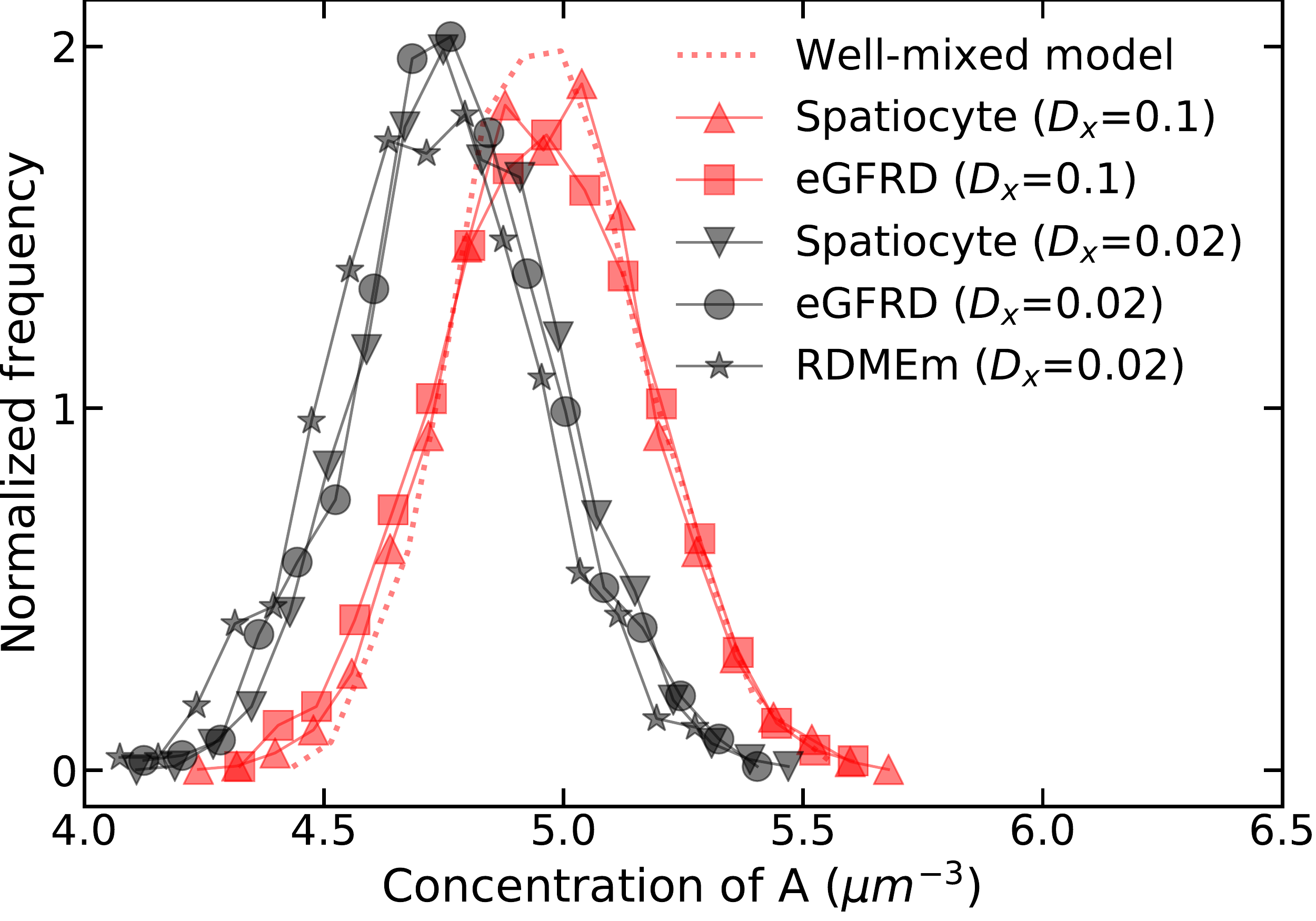}    
  		\caption{}
		\label{fig:dist}
	\end{subfigure}	
	
  \caption{Production-degradation response of A. (a) Time-series profile of $A$ in Eq. \eqref{pdp} simulated with Spatiocyte (using intrinsic rate, $k_a$), eGFRD and RDME. $D_A=D_B=0.1\ \mu \mathrm{m^{2} s^{-1}}$, molecule radius $r_A=r_B=r\in\{0.005,0.02, 0.05\}\ \mu $m. Note that $r$ represents half of the subvolume size in RDME, and the actual molecule radius in Spatiocyte and eGFRD. For comparison, solid line shows the well-mixed model. (b) Mean equilibrium concentration of $A$ from Spatiocyte, eGFRD, RDME and RDME with modified propensity (RDMEm) with $D_A=D_B=D_x\in\{0.01,0.02, 0.1\}\ \mu \mathrm{m^{2} s^{-1}}$. Solid and dashed lines represent expected results according to the well-mixed model and the microscopic theory respectively. (c) Steady-state distribution of $A$ from Spatiocyte and eGFRD with $r=0.05\ \mu $m and $ D_A=D_B=D_x\in\{0.1, 0.02\}\ \mu \mathrm{m^{2} s^{-1}}$. RDMEm simulated with $r=1$ and $D=0.02$ is also shown for comparison. The frequency is normalized such that the sum over the bin is unity. Dotted line represents the well-mixed model simulated using the Gillespie method. Simulation parameters: $k_1=0.1\ \mu \mathrm{m^{-3}s^{-1}}$, $k_2=0.02\ \mu \mathrm{m^{3} s^{-1}}$, $[B]=1\ \mu \mathrm{m^{-3}}$, runs = 700, duration $> 10^4$ s to achieve steady-state, volume = $100\ \mu \mathrm{m^{3}}$ with periodic boundary.}
\label{birthdeath}
\end{figure}

Consider the production and degradation processes of protein $A$ represented by a zero-order production coupled with a second-order degradation: 
\eq{pdp}{\emptyset {\stackrel{k_1}{\xrightarrow{}}}A, \ A+B{\stackrel{k_2}{\xrightarrow{}}}B.}
The concentration of $A$ will go through an initial transient state before settling down at a steady-state equilibrium, $[A]=k_1/(k_2[B])$ that fluctuates according to the Poisson distribution \cite{erban09}.  To confirm if MLM can recapitulate the production-degradation process correctly in 3D space, we have simulated the process with Spatiocyte and compared the outcomes with eGFRD and the well-mixed model. To generate the results of the well-mixed model, we solved the rate equation using an ordinary differential equation (ODE) solver. The time-series of $A$ is shown in Figure \ref{fig:Acurve}, while the equilibrium values are provided in Table \ref{t3}. As evident from the figure and table, Spatiocyte results are all in good agreement with both the well-mixed model and eGFRD. 

Recently, the Spatiocyte scheme was reported to not only fail to reproduce the expected equilibrium value of $A$ but also generate different values depending on the voxel size \cite{Sturrock2016}. In the report, the effective bimolecular rate $k_2$ was used in the calculation of reaction acceptance probability instead of the intrinsic reaction rate constant $k_a$, which inevitably caused the deviation from the well-mixed model (see first row of Table \ref{t3}). As shown in Figure \ref{fig:Acurve} and Table \ref{t3}, there was no discrepancy when the intrinsic rate $k_a$ was used to compute the reaction-acceptance-probability \eqref{pa}. The well-known relation between $k_a$ and $k_2$ is given by Eq. \eqref{keff1}, wherein $k_2$ is represented by $k_{eff}$.
Furthermore, just as in the well-mixed and eGFRD models, the resulting equilibrium concentration from Spatiocyte is also independent of the molecule radius or spatial discretization. Conversely, the RDME method deviated substantially from the well-mixed result when the voxel size is small, which is expected \cite{Fange2010,Hellander2012,isaacson2013}.

\begin{table}[h]
	\caption{Equilibrium concentration of $A$ in Eq. \eqref{pdp} simulated with Spatiocyte and eGFRD at different spatial discretizations. $k_2$ is the effective rate, $k_a$ is the intrinsic rate, $l$ is the voxel size, $K=2^{1/6}L/l$ is the compartment length in number of voxels, while $L$ denotes the actual length \cite{Sturrock2016}. At $l=0.01$, $K=521$; at $l=0.04$, $K=130$; and at $l=0.1$, $K=52$. The well-mixed equilibrium concentration is $5\ \mu \mathrm{m^{-3}}$. Discrepancy (\%) from the well-mixed model is shown in parenthesis. Simulation parameters: production rate, $k_1=0.1\ \mu \mathrm{m^{-3}s^{-1}}$; degradation rate, $k_2=0.02\ \mu \mathrm{m^3 s^{-1}}$; $[B]=1\ \mu \mathrm{m^{-3}}$; volume $=100\ \mu \mathrm{m^3}$; $D_A=D_B=0.1\ \mu \mathrm{m^2 s^{-1}}$; runs = 600.}
\label{t3}
\centering
\begin{tabular}{ l *{4}{l} }
\hline
\hline
Simulation scheme	    	&$l=0.01$		&$l=0.04$ 		&$l=0.1$ \\
\hline
Spatiocyte with $k_2$	    &9.014\ (80.28) &6.023\ (20.46) &5.393\ (7.86) 		\\
Spatiocyte with $k_a$ 		&5.009\ (0.18)  &4.984\ (0.32)  &4.990\ (0.2) \\
eGFRD 						&4.968\ (0.64)  &4.975\ (0.5)   &4.950\ (1) 	    		\\
\hline
\hline
\end{tabular}
\end{table}

The well-mixed model assumes the time scale of diffusion to be always shorter than that of the reactions. As a result, molecules are expected to be uniformly distributed at all times and reactions can take place independent of spatial localization. The well-mixed assumption is valid when describing activation-limited reactions but when they are diffusion-influenced, the position of molecules should be taken into account. We therefore expected some disparity between the well-mixed model and MLM when the production-degradation process is diffusion-influenced. In Figure \ref{fig:Ddpnt}, at smaller diffusion coefficients ($D_x=0.01,0.02$), the equilibrium concentrations are indeed lower with Spatiocyte than with well-mixed model. Spatiocyte behavior is consistent with eGFRD, which also accounts for molecule positions. RDME however, has the same outcomes as the well-mixed model. 

The reduction in equilibrium value when the diffusion coefficient is decreased was previously described by the microscopic theory of Agmon and Szabo \cite{agmon1990}. In contrast to the Collins-Kimball theory, Agmon and Szabo have considered the non-negligible effect of $B$ concentration on the effective reaction rate, especially when the reaction is diffusion-influenced. The slow diffusion of molecules increases the effective contact radius, resulting in higher effective annihilation rate (see Appendix \ref{apdp} for a detailed argument). The output of the production-degradation process according to the microscopic theory is shown in Figure \ref{fig:Ddpnt} as a dashed line that coincides with Spatiocyte and eGFRD, further verifying the MLM theory. Given the same diffusion and macroscopic reaction rates, the change in the Spatiocyte voxel size does not affect the equilibrium behavior (at $r=0.1$ and $r=0.05$ in Figure \ref{fig:Ddpnt}) since the reaction acceptance probability, $P_a$ is adjusted according to the voxel size to obtain the correct macroscopic behavior.

On the other hand, RDME shows large deviation from the expected values at slow diffusion. The inability of conventional rate equation and RDME to correctly capture diffusion-influenced reactions has previously been noted and worked on before \cite{erban09,Fange2010,Hellander2012,isaacson2013,smith2016}. By incorporating the diffusion coefficient into the bimolecular reaction propensity formula (Eq. 26 in \cite{erban09}), the equilibrium concentration of RDME shows a better agreement with the expected values (see RDMEm, $r=1.0$ in Figure \ref{fig:Ddpnt}). However, when the reaction is diffusion-limited ($D_x=0.01,0.02$), unlike MLM, the subvolume size of RDMEm cannot reach the microsopic resolution, $r=0.05$. This is because the size is constrained by a critical value (Eq. 25 in \cite{erban09}) that preserves the well-mixed condition. At $D_x=0.01$ for example, the critical subvolume size is about 13 times the molecule diameter, any size smaller is invalid.

We have also examined the fluctuation of $A$ at equilibrium, as depicted in Figure \ref{fig:dist}. At $D_x=0.1$, the histogram of Spatiocyte matches the distribution curves of eGFRD and the well-mixed model (Gillespie method \cite{gillespie1976}). At much reduced diffusion coefficient ($D_x=0.02$) however, both Spatiocyte and eGFRD shared similar distributions, with the width becoming narrower and the mean value shifting to the left. With the modified propensity function, RDMEm also exhibited similar distribution. The narrow width and the shifted mean are consistent with the characteristics of the Poisson distribution. 

It was reported that MLM would not be able to solve the first-order production-degradation reaction $\emptyset \underset{k_2}{\stackrel{k_1}{\rightleftharpoons}} A$ accurately because of its spatial discretization scheme \cite{Sturrock2016}. When the number of total voxels in the compartment, $N_v$ is less than $k_1/k_2$, the equilibrium concentration deviates from the well-mixed model. This deviation however, is a direct consequence of the volume exclusion property of MLM. Since each voxel can only occupy a single molecule, there would be an insufficient number of vacant voxels to accommodate new molecules when the degradation rate is not sufficiently fast. The maximum occupancy on HCP lattice simply reflects the maximum physical occupancy of voxel-sized molecules in the compartment because the HCP arrangement packs the highest density of sphere voxels \cite{szpiro2003kepler}. Just as in the cellular compartment, no more molecules can be added into the system when the number of generated molecules exceeds available free space. Moreover, since only about 34 \% of the cell volume is occupied by macromolecules \cite{zimmerman1991estimation}, it is also an unlikely scenario to fully occupy the voxels of HCP lattice with macromolecules. With the multi-algorithm implementation of Spatiocyte \cite{arjunan2017multi}, we can use the Gillespie's Next-Reaction method \cite{gibson2000} to simulate small molecules that are in large abundance and are homogeneously distributed. In this case, the equilibrium result is independent of spatial discretization since the method assumes the well-mixed condition.

\subsubsection{Dual phosphorylation-dephosphorylation cycle} 
\begin{figure}[!h]
  \centering
    \begin{subfigure}[t]{0.49\textwidth}
  	\centering
    \includegraphics[width=\textwidth]{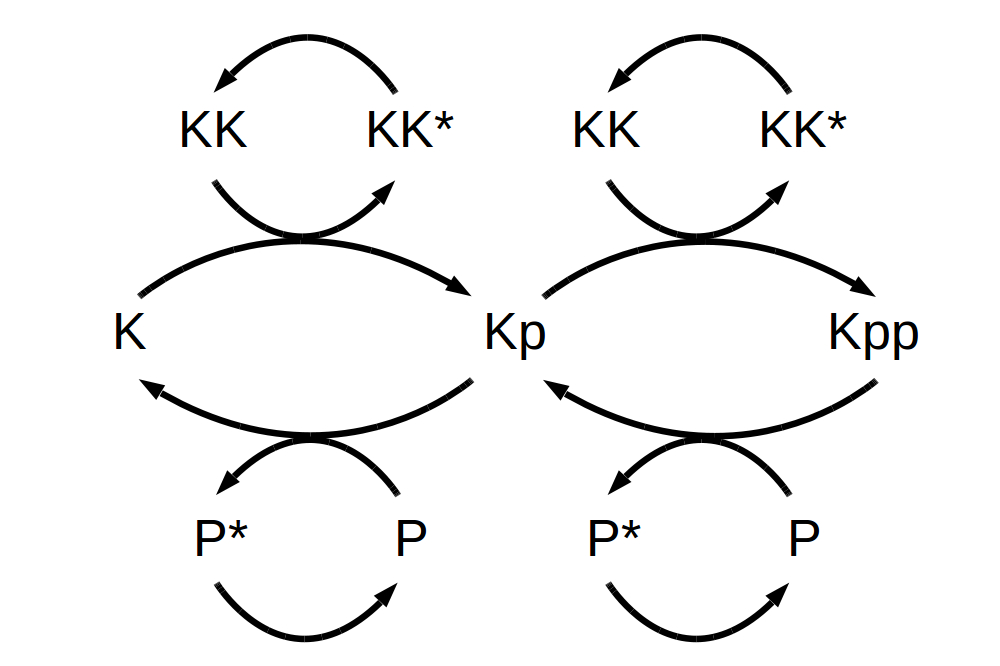}    
  	\caption{}
	\label{fig:rxnmodel}
	\end{subfigure}
	\hfill
	\begin{subfigure}[t]{0.49\textwidth}
   	\centering	  
    \includegraphics[width=\textwidth]{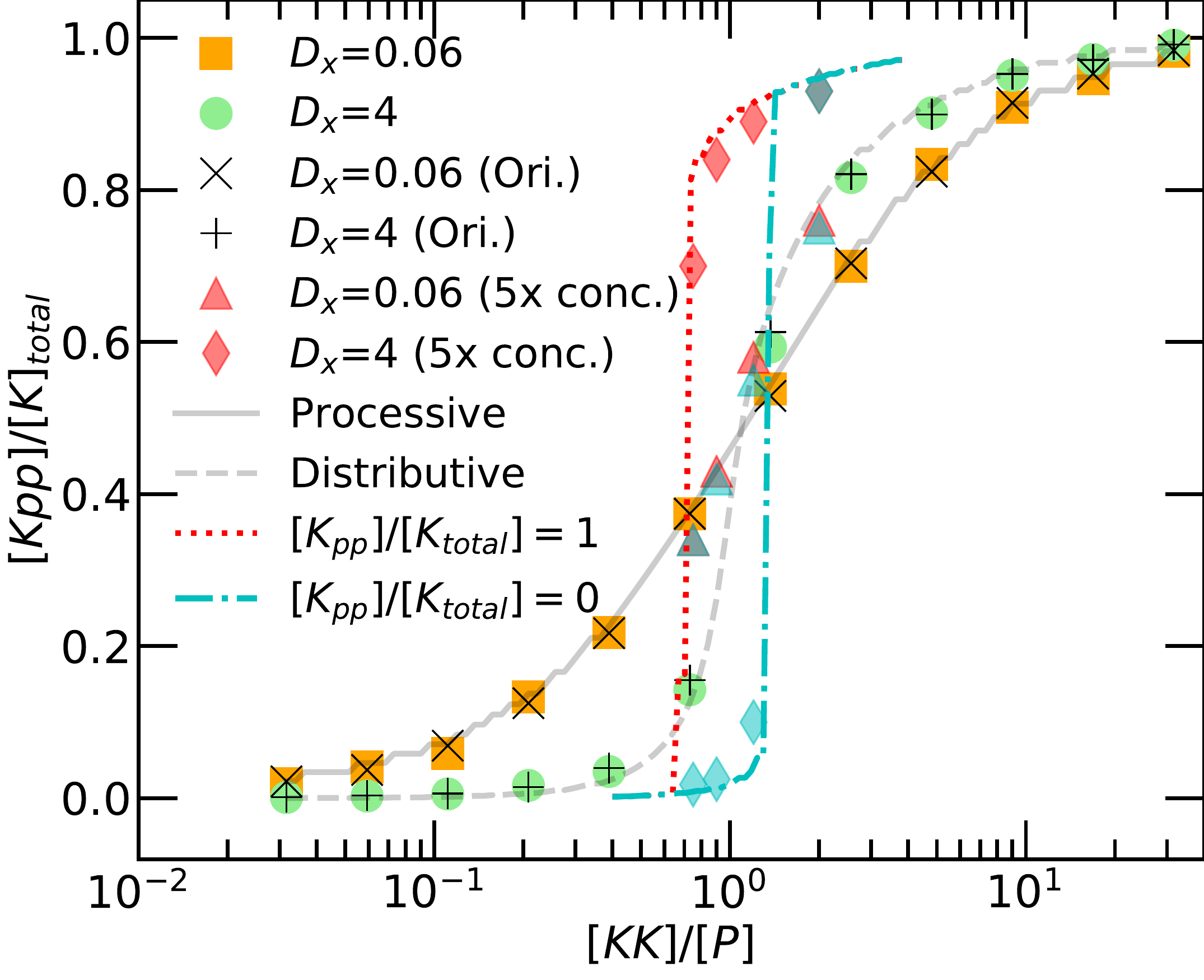}    
  	\caption{}
	\label{fig:response}   	
	\end{subfigure}
  \caption{Effects of rebinding in dual phosphorylation cycle. (a) Reaction model showing MAPK (K) is first activated into Kp and then Kpp by MAPKK (KK) in two phosphorylation steps. Kpp is also deactivated by phosphatase (P) in two dephosphorylation steps to become K again. Enzymes KK and P become inactive immediately after reacting with their respective substrates and then relax back to the active state after some delay $\tau_{rel}$. (b) Fraction of Kpp in response to MAPKK/phosphatase ratio at steady-state. Circle and square markers denote simulation result using Spatiocyte with $D_x=4\ \mu \mathrm{m^2s^{-1}}$ and $D_x=0.06\ \mu \mathrm{m^2s^{-1}}$, respectively. Dashed and solid lines represent distributive and processive mechanism models, respectively. Cross and plus markers show the results from the original Spatiocyte scheme, wherein the voxel and molecule sizes are exactly the same. We used a short reactivation time, $\tau_{rel} = 1\ \mu $s, relative to $t_d$ (for comparison $t_d \approx 1\ \mu $s when $D_x=4\ \mu \mathrm{m^{2} s^{-1}}$, $t_d\ \approx 70\ \mu $s when $D_x=0.06\ \mu \mathrm{m^{2} s^{-1}}$) with the total number of substrates, $\mathrm{K_{total}}=120$. Hysteresis responses from mean-field distributive model with five-fold substrate concentration ($\mathrm{K_{total}}=600$) are indicated by dotted and dash-dotted lines with initial conditions $\mathrm{[K_{pp}]/[K]_{total}}=1$ and $[\mathrm{K_{pp}]/[K]_{total}}=0$, respectively. Diamond and triangle markers represent Spatiocyte responses with five-fold substrate concentration at the indicated diffusion coefficient, $D_x$. Simulation parameters: molecule size $l=0.0025\times1.0209\ \mu $m, diffusion coefficient $D_x$, $\mathrm{[KK]+[P]}=60$, duration = 200 s, volume = $1\ \mu \mathrm{m^3}$ with periodic boundary. 
  }
\label{mapk}
\end{figure}

In mean-field models, the spatio-temporal correlation of microscopic rebinding events is not resolved explicitly because the correlation usually does not cause a significant impact on the dynamics at the macroscopic scale. One case where the correlation does influence the macroscopic response is the dual phosphorylation-dephosphorylation cycle of the MAPK cascade \cite{chang2001mammalian,ferrell1997,aoki2011}, shown in Figure \ref{fig:rxnmodel}. The substrate MAPK (K in Figure \ref{fig:rxnmodel}) is phosphorylated in a two-step process by the MAPK kinase (KK) and dephosphorylated by a phosphatase P. The phosphorylation and dephosphorylation processes proceed according to the Michaelis-Menten kinetics and exhibit distributive property \cite{ferrell1997}, wherein the enzymes must unbind from the substrate before they can rebind and modify the second site. Upon phosphorylation or dephosphorylation, the respective enzymes are inactivated (denoted as KK* and P*), and reactivated (KK or P) after some time $\tau_{rel}$. When the reactivation time is short and the enzyme-substrate reaction is diffusion-limited, the newly dissociated enzyme and substrate are close enough to rebind instead of escaping into the bulk. These microscopic rebinding events alter the response sensitivity of the phosphorylation state as shown by Takahashi et al. \cite{takahashi2010} using eGFRD. Processive behavior caused by rebindings of the same enzyme results in higher overall phosphorylation rate than the distributive case where the dissociated molecules can escape rebinding \cite{ferrell1997,aoki2011}. Such microscopic spatio-temporal correlation has been shown to change the response sensitivity of the phosphorylation state, which can cause the subsequent removal of ultra-sensitivity or bi-stability in the system \cite{elf2004,takahashi2010}.

Rebinding events taking place within very short time scales are difficult to be captured by RDME because of the fine spatial resolution required. To test whether MLM can resolve such events faithfully, we use Spatiocyte to model the dual phosphorylation cycle with the same parameters from \cite{takahashi2010}. Distributive and processive models are represented by Eqs. 1-5 of \cite{takahashi2010}, and were solved using ODE solver. Figure \ref{fig:response} displays the steady-state response curves of Spatiocyte and reference theoretical models. Note that since the reactivation time $\tau_{rel}$ is equal to or less than the diffusion time step $t_d$ (given in Figure \ref{fig:response}), the molecules can rebind soon after dissociation. The simulation result coincides very well with the switch-like response curve of the distributive model at fast diffusion ($D_x=4\ \mu \mathrm{m^{2} s^{-1}}$), whereas at much slower diffusion ($D_x=0.06\ \mu \mathrm{m^{2} s^{-1}}$), it converges to the graded response curve of the processive model. The influence of diffusion on the response curve can be understood through the rebinding events. When diffusion is slow, reactions become more diffusion-limited and rebinding occurs at higher frequency. The ensuing processive-like mechanism then leads to the loss of the switch-like response curve. Conversely, in the limit of fast diffusion as assumed in the mean-field model, a sharper switch-like response curve is recovered because of fewer rebindings.

The parameter ranges examined so far have a stable steady-state as demonstrated by the response curves in Figure \ref{fig:response}. When the total concentration of the substrate is increased five-fold, the mean-field theory generates hysteresis, shown by the dotted and dash-dotted lines. The dotted line represents the response when initialized with $\mathrm{[K_{pp}]/[K]_{total}}=1$, whereas the dash-dotted line has the initial condition $\mathrm{[K_{pp}]/[K]_{total}}=0$. MLM produced similar responses when the diffusion is fast $(D_x=4)$ (diamond markers in Figure \ref{fig:response}). However, as diffusion slowed down to $D_x=0.06$, the bistability is lost (triangle markers). Bistable states appear when the diffusion is fast and the substrate concentration relative to enzyme is high. For example, at the initial state when almost all substrates are in the unphosphorylated form, most kinase will be bound to the substrates rapidly. Hence, a substrate that has been phosphorylated once is more likely to be dephosphorylated by free phosphatase than to be phosphorylated the second time by scarce and fast diffusing kinase. The inverse situation where all substrates are in the phosphorylated form would also respond similarly to phosphatase. On the other hand, when diffusion is slow, the kinase activity becomes processive because of the high rebinding probability. As a result, molecules are more likely to be phosphorylated or dephosphorylated consecutively before they could be disrupted by antagonistic enzymes from the bulk. This example highlights how local spatio-temporal correlation can change the binding behavior and results in a different global response than the one predicted by the mean-field model.

As a side remark, in the original Spatiocyte scheme \cite{Arjunan2010}, the voxel adopts the size of the diffusing molecules. However, as we found in the Methods section, the voxel needs to be about 2\% larger than the molecule size \eqref{lfactor} for the total rebinding probability and the effective rate constant to be exactly the same as in the continuum-based theory. Despite the 2\% difference in voxel sizes, both new and original schemes displayed very good fit with the expected dual phosphorylation cycle response curves in Figure \ref{fig:response}. To be fully consistent with the continuum-based theory however, the size should be set according to Eq. \eqref{lfactor}. The voxel size is not hard-coded to be the same as the molecule size and can be easily specified in the Spatiocyte model file \cite{arjunan2017multi}.

\subsubsection{Effects of excluded volume on bimolecular reaction}
Excluded volume in the cell arising from crowded obstacles such as macromolecules, Golgi apparatus or cytoskeletal elements can cause anomalous diffusion of reacting molecules \cite{Saxton2002,vilaseca2011}. Anomalous diffusion has been shown to generate non-classical reaction kinetics on 2D  \cite{Berry2002a,Schnell2004} and 3D lattices \cite{pitulice2014}. Here, we use MLM on HCP lattice to examine the effects of volume exclusion on the bimolecular reaction $E+S\xrightarrow{}\emptyset $ in the presence of uniformly distributed immobile obstacles. $E$ and $S$ have the radius $5\ \mathrm{nm}$ and diffusion coefficient, $D_0=1\ \mu \mathrm{m^2s^{-1}}$. Hence, $D_0$ is the diffusion coefficient in non-crowded dilute condition. Bimolecular intrinsic reaction rate constant $k_a=10k_D$ is chosen such that the reaction is diffusion-limited. Excluded volume is quantified by the lattice occupancy of the obstacles, $\phi=N_o/N_v$, where $N_o$ and $N_v$ are the numbers of obstacles and total voxels, respectively. Simulation is carried out in a periodic cubic compartment with length $L=1\ \mu \mathrm{m}$ for a duration of $1000t_d$. Reactants have dilute concentrations, $[S]=5[E]=0.001N_v$ and are placed randomly at the beginning of simulation. 

\begin{figure}[!h]
  \centering
    \begin{subfigure}[t]{0.49\textwidth}
	  	\centering
    		\includegraphics[width=\textwidth]{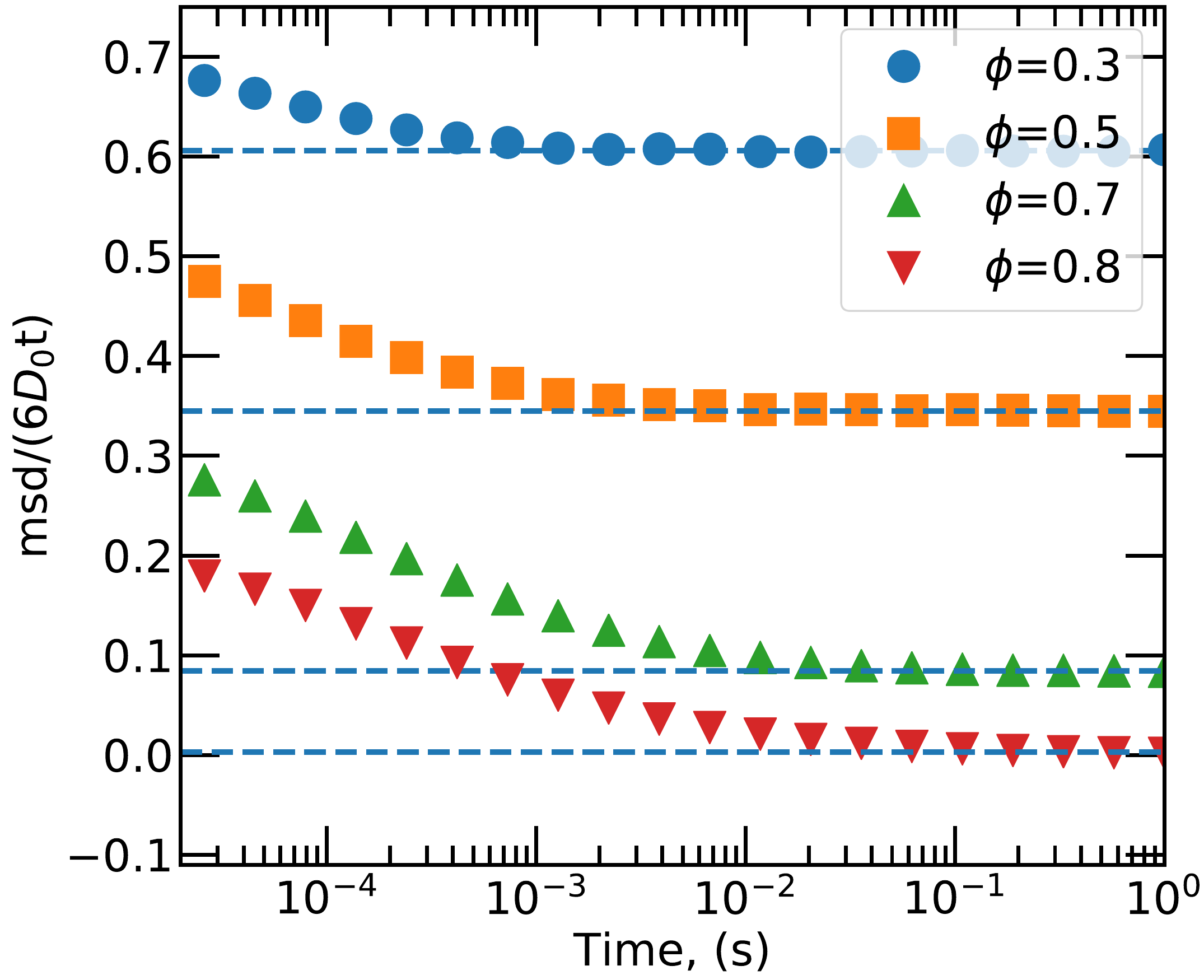}    
	    \caption{}
		\label{fig:crowdingdt}   	
	\end{subfigure}
	\hfill
	\begin{subfigure}[t]{0.49\textwidth}
	   	\centering	  
    		\includegraphics[width=\textwidth]{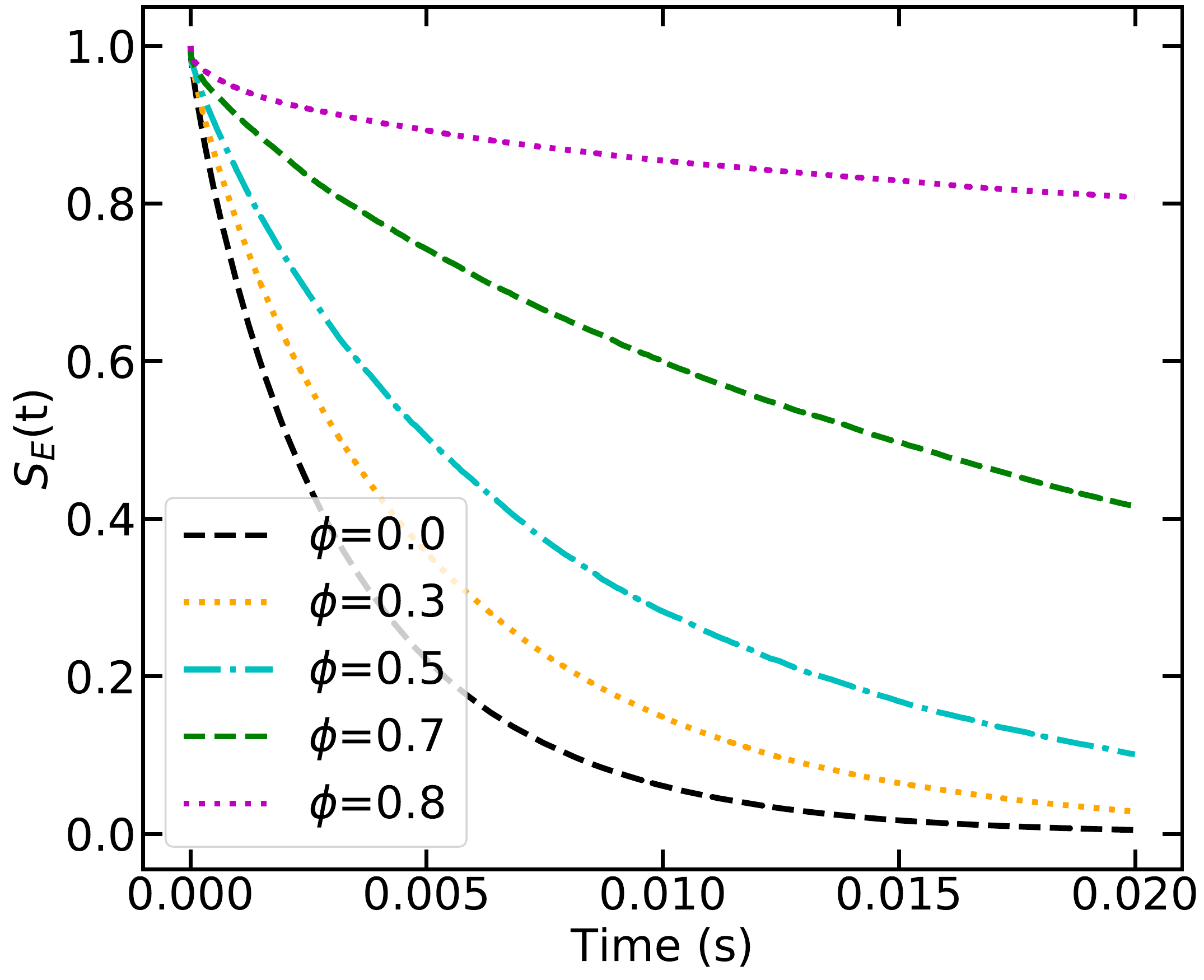}    
	    \caption{}
		\label{fig:crowdingst}
	\end{subfigure}
	\begin{subfigure}[t]{0.49\textwidth}
	   	\centering	  
    		\includegraphics[width=\textwidth]{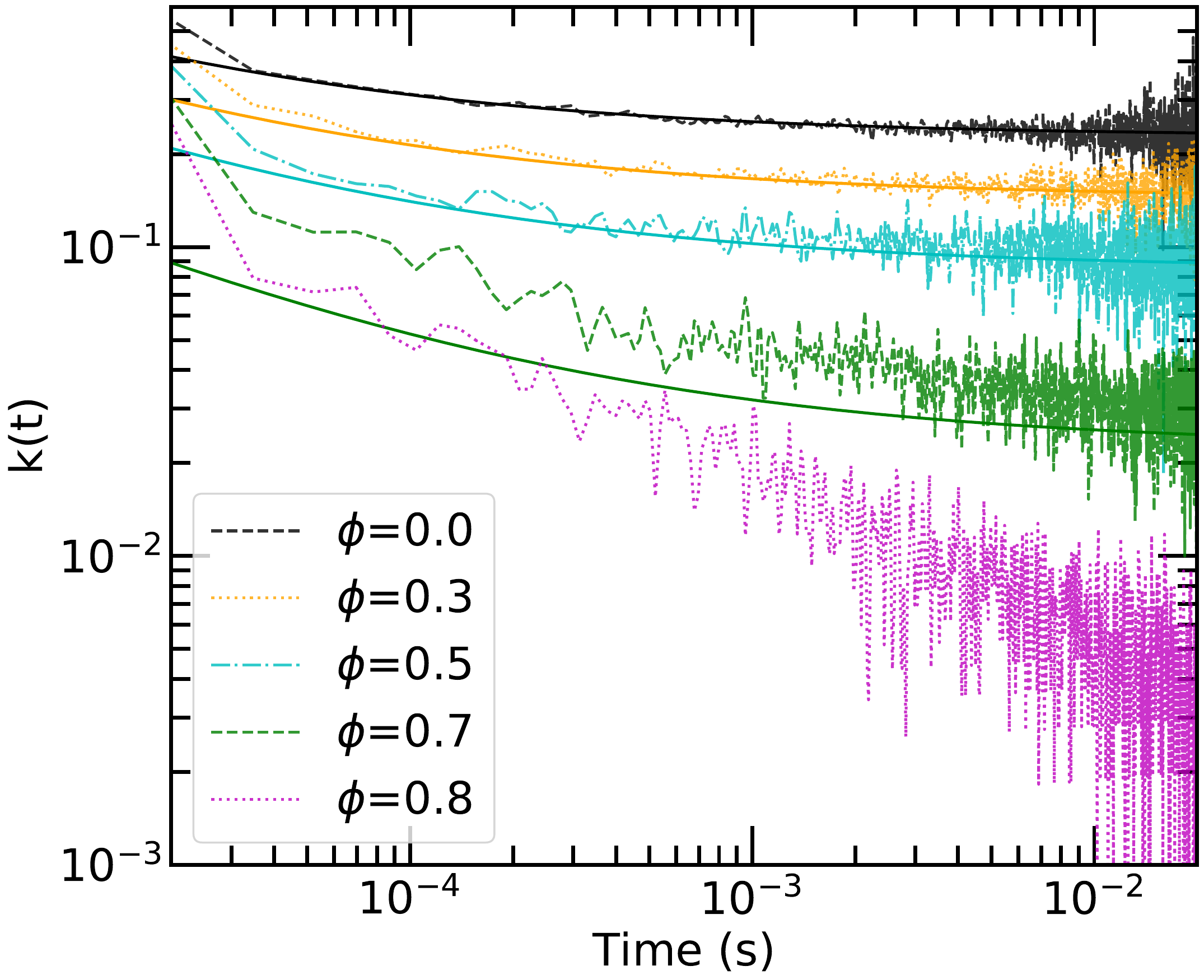}    
	    \caption{}
		\label{fig:crowdingkt}   	
	\end{subfigure}	
  \caption{Diffusion and bimolecular reaction kinetics in crowded compartment. (a) Time-dependent diffusion coefficient of tracer molecules in the presence of immobile obstacles at volume occupancy, $\phi$. The diffusion coefficient at a time point is determined from the mean-squared displacement of simulated particle trajectories. Dashed lines denote the diffusion coefficient at long-time as predicted by $D'=D_0(1-\phi/\phi_p)$. (b) Survival probability of $E$ in $E+S\xrightarrow{}\emptyset $ at $\phi$. (c) The corresponding time-dependent reaction rates (dashed lines) at $\phi$. Solid lines represent Collins-Kimball theory with the long-time diffusion coefficient calculated in (a). Simulation parameters: compartment volume = $(1\ \mu \mathrm{m})^3$ with periodic boundary, $R=0.01\ \mu \mathrm{m}$, $l=0.01\times1.0209\ \mu \mathrm{m}$, $D_E=D_S=1\ \mu \mathrm{m^{2} s^{-1}}$, $k_a=10k_D$, $[S]=5[E]=0.001N_v$, duration = 0.02 s.
  }
\label{crowding}
\end{figure}

We first consider the effects of immobile obstacles on diffusing molecules. We calculate the time-dependent diffusion coefficient from the mean-squared displacement of simulated particle trajectories. The time-dependent diffusion coefficient in Figure \ref{fig:crowdingdt} indicates that the diffusion is anomalous at short times and normal at long times. The crossover time from anomalous to normal diffusion depends on the volume occupancy. The reduced long-time diffusion coefficient is well-described by \cite{saxton1989lateral,vilaseca2011},
\eq{slowd}{
D'=D_0(1-\phi/\phi_p) 
,}
where $\phi_p\approx 0.77$ is the percolation threshold for HCP lattice. We confirmed that the long-time diffusion coefficients obtained for $\phi$ in Figure \ref{fig:crowdingdt} (dashed lines) are consistent with $D'$ in Eq. \eqref{slowd}.

Figure \ref{fig:crowdingst} shows that the survival probability of $E$ decays slower when the volume occupancy, $\phi$ is increased. From the survival probability, we can calculate the rate coefficient according to Eq. \eqref{ktcal} to obtain the kinetics. We replaced the constant concentration term $[B]$ \eqref{ktcal} with the time varying term $[E](t)$ in the equation. For the dilute case ($\phi=0$) in Figure \ref{fig:crowdingkt}, there is a good agreement for the simulated $k(t)$ with the Collins-Kimball rate coefficient \eqref{ktck}. As $\phi$ increases to 0.3 and 0.5, the overall reaction rate decreases, and thus progressively diverges from the Collins-Kimball rate. Despite the discrepancy, the rates can still conform to the Collins-Kimball theory when the long-time diffusion coefficient \eqref{slowd} is used.

As the volume occupancy approaches the percolation threshold (Figure \ref{fig:crowdingkt}, $\phi=0.7$), the kinetics begins to deviate from the Collins-Kimball theory. The deviation is strongest at $\phi=0.8$, which is beyond the percolation threshold. Note that at lower volume occupany ($\phi=0.3,0.5$), the anomalous to normal diffusion crossover time in Figure \ref{fig:crowdingdt} is faster than the observation time in Figure \ref{fig:crowdingkt}. Here, the kinetics is well described by the long-time effective diffusion coefficient. However, when the crossover time is comparable to the observation time because of the increased volume occupancy (Figure \ref{fig:crowdingdt}, $\phi=0.7$), the effects of anomalous diffusion is visible in the kinetics (Figure \ref{fig:crowdingkt}, $\phi=0.7$). At above the percolation threshold ($\phi=0.8$), anomalous diffusion does not crossover to normal diffusion. As a result, the long-time diffusion coefficient eventually decays to zero. In these highly crowded cases, the Collins-Kimball theory fails to describe the kinetics.

Grima and Schnell \cite{Grima2006} have shown that reaction kinetics, either classical or non-classical, is not determined by the heterogeneity of the accessible space but rather by the reaction probability and the initial condition. In the Smoluchowski and Collins-Kimball framework, reaction follows classical kinetics when it is activation-limited ($k_a/k_D\ll 1$) but non-classical kinetics is observed when it is diffusion-influenced ($k_a/k_D\gg 1$). The non-classical behavior in the latter is well-described by Eq. \eqref{ktck} using microscopic parameters. The corresponding long-time behavior up to the second order term scales according to Eq. \eqref{klarget}, which has the same general form of the Zip-Mandelbrot equation proposed by Schnell and Turner \cite{Schnell2004,pitulice2014}. The Zip-Mandelbrot equation is valid for long-time kinetics whereas the Collins-Kimball rate \eqref{ktck} describes the kinetics for all time ranges. 

Here, we have studied the kinetics of bimolecular reaction in the presence of immobile obstacles with MLM. When the total volume occupied by obstacles is much smaller than the percolation threshold and the observation time scale is longer than the anomalous to normal diffusion crossover time, the kinetics is still reproducible with the Collins-Kimball theory and Eq. \eqref{slowd}. However, it deviates from the theory when the volume occupancy nears or crosses the percolation threshold, wherein anomalous diffusion dominates and the diffusion coefficient approaches zero at the long-time limit. Therefore, to better describe the non-classical kinetics analytically, we should incorporate the anomalous diffusion induced by fractal medium into the theory either phenomenologically \cite{kopelman1988fractal,Schnell2004,pitulice2014} or by extending the Smoluchowski and Collins-Kimball framework using a generalized diffusion equation \cite{barzykin1993,sung2001green}. 

\section{Conclusions}

In contrast to macroscopic and mesoscopic approaches, particle-based methods have the advantage to directly link microscopic parameters to the observed RD behavior, thus providing insights about the underlying mechanisms of the system. MLM shares this same advantage, but with reduced computational costs owing to its fixed step lengths and voxel-based collision detection algorithm. The reduction in computational costs allows MLM to not only simulate non-dilute and crowded intracellular conditions \cite{andrews2015simulating} but also track individual molecules on large eukaryotic cells \cite{watabe2015computational} and simulate membrane protein clustering in whole red blood cells \cite{Shimo2015}. 

Recently, Grima and colleagues developed a method called vRDME that incorporates volume exclusion into RDME \cite{cianci2016,smith2016}. The method can approximate the continuum model very well by matching the steady-state rate constants of both models. We note that vRDME is a type of MLM since each voxel can occupy a molecule and bimolecular reactions occur by colliding reactants. In contrast to vRDME, our work here employs random walk theory and particle-pair formalism to describe bimolecular reactions. Notably, both the effective rate constant and the total rebinding probability on lattice are matched to the corresponding continuum expressions to determine the correct reaction acceptance probability and voxel size. 

Contrary to the original assumption of Spatiocyte \cite{Arjunan2010}, the voxel should be larger than the molecule size (by about $2\%$ for HCP lattice) to be quantitatively accurate. Numerical simulations showed that both the effective rate constant and the asymptotic time-dependent behavior have good agreements with the Collins-Kimball theory in activation- and diffusion-limited cases. MLM also displayed very good consistencies with eGFRD when simulating actual biochemical systems such as protein production-degradation and the dual phosphorylation cycle. Although MLM is analyzed based on the HCP lattice in this work, the theoretical framework is also applicable for other lattice arrangements such as cubic lattice, by simply updating the lattice density and the return probability $F(1)$ (see Appendix \ref{alir} and \ref{avs}).

Despite achieving the same total rebinding probability as the Collins-Kimball theory, the time-dependent behavior of MLM at time scales shorter than $t_d$ is different than that theory (Figure \ref{rebinding}). One potential solution to obtaining the same behavior at such fine time scales is to make the voxel size smaller than the molecule, similar to the SVTA approach \cite{Gillespie2014}. This would reduce $t_d$ but increase the cost of computation significantly because of the finer time steps and the higher number of collision checks required.

MLM captures the effects of excluded volume naturally but comparing on-lattice behavior with continuum is not straightforward since the influence of volume exclusion and the resulting reaction kinetics vary according to the lattice arrangement \cite{Grima2006,meinecke2016}. Moreover, since all diffusing species in this work have the same molecule size, it is not possible to replicate the effects of relative size of interacting molecules. To minimize such lattice artifacts and to better approximate off-lattice volume exclusion, we can improve the size representation of each molecule on lattice by occupying multiple voxels as in the SVTA approach or by employing a hybridized on- and off-lattice approach. Higher spatial resolution of molecules would generate more realistic diffusion behavior in a crowded environment. Alternatively, we can introduce a density-dependent hopping rate as adopted by two previous RDME methods \cite{grima2007mesoscopic,cianci2017}. 

Realistic simulation of intracellular reaction-diffusion processes should also incorporate the influence of inter-molecular potentials such as van der Waals and hydrodynamic forces. By employing contact interactions on lattice as proposed by Fernando et al. \cite{fernando2010} or the SVTA approach with interaction potentials \cite{Gillespie2014}, it may be possible to incorporate the above forces in MLM. The theoretical framework presented in this work serves as a building block for further development and integration of MLM-based algorithms.

\section{Author Contributions}
W.-X.C., K.K., K.T. and S.N.V.A. designed research; W.-X.C. performed research;  W.-X.C., K.K., M.W., S.V.M., and S.N.V.A. analyzed data; and W.-X.C., K.K. and S.N.V.A wrote the manuscript. All authors read and commented on the manuscript.

\section{Acknowledgements}
We thank Kozo Nishida for technical advice and support and Kylius Wilkins for critical reading of the manuscript. We thank Ramon Grima for providing the Matlab source of fast Brownian dynamics (point particle version) and Steven Andrews for his help with Smoldyn models used in the performance benchmarks. W.-X.C. acknowledges RIKEN for supporting his doctoral research as an International Program Associate.

\appendixtitleon
\numberwithin{equation}{section}
\begin{appendices}
\section{Rebinding probability distribution}\label{arpd}
The rebinding probability distribution is defined as (Eq. 3.10 in \cite{naqvi1980} and Eq. S27 in \cite{takahashi2010})
\eq{RP}{p_{reb}(R,t;R,0)=k_a p(R,t;R,0),}
where $p(r,t;r_0,0)$ is the Green's function in the diffusion equation:
\eq{}{\frac{\partial p(r,t;r_0,0)}{\partial t}=D\nabla^2 p(r,t;r_0,0),}
subjected to initial condition
\eq{}{p(r,0)=\frac{\delta(r-r_0)}{4\pi r^2},}
and boundary conditions such that
\eq{}{p(r,t)\to 0 \text{ as }r\to \infty.}
and 
\eq{}{4\pi R^2D\left.\frac{\partial p(r,t;r_0,0)}{\partial r} \right|_{r=R}=k_a p(R,t;r_0,0).}
The latter condition is known as the radiation boundary condition.
The Green's function $p(r,t;r_0,0)$ has been solved in (p. 368 in \cite{carslaw1959}) to be
\meq{}{p(r,t;r_0,0)&=\frac{1}{8\pi rr_0}\frac{1}{\sqrt{\pi Dt}}\left[\exp{[-(r-r_0)^2/4Dt]}+\exp{[-(r+r_0-2R)^2/4Dt]}\right.\\
&\left. -2B\sqrt{\pi D t}\exp{\left[B^2Dt+B(r+r_0-2R)\right]}\text{erfc}\left(B\sqrt{Dt}\right) \right],}
where $B=(1+k_a/k_D)/R$.
\\For $r=r_0=R$, we thus have
\meq{GF}{p(R,t;R,0)&=\frac{1}{4\pi R^2}\frac{1}{\sqrt{\pi Dt}}\left[1-B\sqrt{\pi D t}\exp{\left(B^2Dt\right)}\text{erfc}\left(B\sqrt{Dt}\right) \right].\\}
Finally by substituting Eq. \eqref{GF} into Eq. \eqref{RP}, we obtain the probability distribution
\eq{}{p_{reb}(R,t;R,0)=\left(\frac{k_a}{4\pi R^3}\right) \left(\frac{k_a}{k_D}+1 \right)\left(\frac{1}{\sqrt{\pi\tau}}-\exp(\tau)\text{erfc}(\sqrt{\tau}) \right),}
where $\tau=tD(1+k_a/k_D)^2/{R^2}$.

\section{Lattice initial rate}\label{alir}
Here we provide the derivation of the lattice initial rate, which was done previously in \cite{Arjunan2010}. Given two reacting species $A$ and $B$, in which $A$ are stationary and $B$ are diffusing. The initial rate constant at time step $t'$, can be estimated using the rate equation as
\eq{}{k_a'=\frac{\Delta N_c V}{N_AN_Bt'},}
where $N_i$ denotes the number of molecules of species $i$, $\Delta N_c$ denotes the change in $N_c$ and $V$ is the compartment volume. The number of successful reactions in a single step $t'$ can be crudely estimated as $\Delta N_C=ZP_a'$ where $Z=N_BN_A/N_v$ is the average number of encounter, $N_v=\sqrt{2}V/l^3$ is the total number of voxels in a compartment volume $V$, and $P_a'=P_a\alpha$ is the actual reaction acceptance probability during the encounter.

For the activation-limited scheme, where $t'=t_d$ and $P_a'=P_a$, 
The initial reaction rate is then given by
\meq{PA}{
k_a'&=\frac{P_a'l^3}{\sqrt{2}t'}\\
&=\frac{P_al^3}{\sqrt{2}t_d}\\
&=3\sqrt{2}P_aDl,
}
Note that $D$ is the sum of diffusion coefficients of the reacting pair, $D_A+D_B$.
\\Similary, for the diffusion-influenced scheme, where $t'=t_d\alpha$ and $P_a'=P_a\alpha$, we have
\meq{}{
k_a'&=\frac{\alpha P_al^3}{\sqrt{2}\alpha t_d}\\
&=3\sqrt{2}P_aDl.
}
Also note that the physical dimension of $k_a'$ satisfies $cm^3s^{-1}$.
\\The above derivation for HCP lattice can be generalized to other lattice arrangements:
\eq{kadensity}{k_a'=\frac{\pi P_aDl}{d},}
where $d$ is the packing density of the lattice (e.g. $d=\pi/6$ for the simple cubic lattice).

\section{Voxel size}\label{avs}
As shown in main text, in order to match the MLM with the continuum-based model, the voxel size of HCP lattice has to be chosen such that 
\eq{vs}{
l=\frac{4\pi R}{ 3\sqrt{2}\left(\frac{1}{F(1)}-1\right)},
}
where $R$ is the molecule size and $F(1)\approx 0.256318$ (p. 153 in \cite{hughes1995}) is the total return probability on HCP lattice. 
\\More generally, the voxel length of any regular lattice arrangement follows that
\eq{}{l=\frac{4d}{\left. \frac{1}{F(1)}-1 \right.}R.
}
For example, for the simple cubic lattice we have the voxel length: 
\eq{}{l=\frac{4\pi/6}{\frac{1}{0.340537}-1}R=1.081515R,
}
about $8\%$ larger than the molecule size ($F(1)$ for the simple cubic lattice is given in p. 153 of \cite{hughes1995}).
\section{First-passage time distribution on HCP lattice}\label{aftd}
For $n\in \mathbb{N}$, we define $P_n(s_a|s_b)$ as the voxel occupation probability from $s_b$ to $s_a$, that is, the probability of being at voxel $s_a$ after $n$ steps, given that the walk started at voxel $s_b$; $F_n(s_a|s_b)$ as the first-passage time distribution from $s_b$ to $s_a$, that is the probability of arriving at $s_b$ for the first time on the $n$th step, given that the walk started at site $s_a$; $s_0$ as the origin voxel, $s_1$ as the element of the set of immediate neighboring voxels of $s_0$, and $s_2$ as the element of the set of the second nearest neighbor voxels of $s_0$.

The probability generating function of $F_n(s_0|s_0)$ and $P_n(s_0|s_0)$ is related through (Eq. I.18 in \cite{montroll1965random})
\eq{FP}{F(s_0|s_0;z)=\sum_{n=0}^\infty F_n(s_0|s_0)z^n=1-\frac{1}{P(s_0|s_0;z)},}
where $P(s_0|s_0;z)=\sum_{n=0}^\infty P_n(s_0|s_0)z^n$ is the lattice Green's function for the face-centered cubic (FCC) lattice as defined in Eqs.(2.6)-(2.9) of \cite{joyce98}:
\eq{}{P(s_0|s_0;z)=\left[\frac{2(1+3\xi^2)}{\pi(1-\xi)(1+3\xi)} \right]^2K(k_+)K(k_-),}
\eq{}{k_+^2=\frac{16\xi}{(1-\xi)(1+3\xi)^3},}
\eq{}{k_-^2=\frac{16\xi^3}{(1-\xi)^3(1+3\xi)},}
\eq{}{\xi=\frac{-1+\sqrt{1+z/3}}{1+\sqrt{1-z}},}
wherein $K$ is the complete elliptic integral of the first kind.
\\For the convenience of calculation, the voxel occupation probability is given as \cite{joyce98}
\eq{}{
P_n(s_0|s_0)=\frac{1}{12^n}\sum_{j=0}^n\binom{n}{j}(-4)^{n-j}b_j,\text{ for }\ j \in \mathbb{N}
}
where 
\eq{}{b_j=\sum_{k=0}^j\binom{j}{k}^2\binom{2k}{k}\binom{2j-2k}{j-k}.} 
The first-passage time distribution is related to the voxel occupation probability recursively via:
\eq{}{F_n(s_0|s_0)=P_n(s_0|s_0)-\sum_{j=1}^{n-1}P_{n-j}(s_0|s_0)F_j(s_0|s_0),\text{ for }\ j \in \mathbb{Z}^+.}
\subsection{Activation-limited case \texorpdfstring{($k_a\ll k_D$, $\alpha=1$)}{Lg}}\label{aalc}
For $P_a=1$, the rebinding-time probability distribution $F_n(s_0|s_1)$ is equivalent to the first-passage time distribution $F_{n+1}(s_0|s_0)$ as mentioned in the main text. 
\\ Whereas for $P_a<1$, the rebinding-time probability distribution is given by
\eq{}{H_n(s_0|s_1)= P_a F^1_{n+1}(s_0|s_0)+P_a(1-P_a) F^2_{n+2}(s_0|s_0)+P_a(1-P_a)^2 F^3_{n+3}(s_0|s_0)+...,}
wherein $F^j_n(s_0|s_0)$ is the probability of reaching the origin for the $j$th time at $n$th step (I.1.9 in \cite{montroll1965random}):
\eq{fnj}{F_n^j(s_0|s_0)=\sum_{i=1}^{n}F_{n-i}^{j-1}(s_0|s_0)F_i(s_0|s_0),\text{ for }\ j \in \mathbb{Z}^+,}
with $F_n^1(s_0|s_0)=F_n(s_0|s_0)$.
\\With Eq. \eqref{fnj} we can obtain $H_n(s_0|s_1)$ recursively via
\eq{hnr}{H_n(s_0|s_1) = P_a \sum_{j=1}^{n} F_{n+j}^{j}(s_0|s_0)(1-P_a)^{j-1},\text{ for }\ j \in \mathbb{Z}^+, \ n \in \mathbb{N}.}
\\The generating function of $H_n(s_0|s_1)$ is related to the generating function of $F_n(s_0|s_0)$:
\meq{hs0s1}{
H(s_0|s_1;z) &=\sum_{n=0}^\infty H_n(s_0|s_1)z^n\\
&=P_a\sum_{n=0}^\infty \sum_{j=1}^\infty F_{n+j}^j(s_0|s_0)\ (1-P_a)^{j-1}z^n\\
&=P_a\sum_{j=1}^\infty (1-P_a)^{j-1}z^{-j} \sum_{n=0}^\infty F_{n+j}^j(s_0|s_0)\ z^{n+j}\\
&=P_a\sum_{j=1}^\infty (1-P_a)^{j-1}z^{-j} \sum_{n=0}^\infty F_{n}^j(s_0|s_0)\ z^{n}\\
}
where in the last step we have $\sum_{k=1}^{j-1} F_{k}^j(s_0|s_0)\ z^{n}=0$ since for all $k$ such that $k< j-1$, the return probability is zero. \\
Using (Eq. I.20 in \cite{montroll1965random}):
\eq{}{\sum_{n=0}^\infty F_{n}^j(s_0|s_0)\ z^{n}=F(s_0|s_0;z)^j,}
in Eq. \eqref{hs0s1} we then have:
\meq{}{H(s_0|s_1;z)&=P_a\sum_{j=1}^\infty (1-P_a)^{j-1}z^{-j} F(s_0|s_0;z)^j\\
&=\frac{P_aF(s_0|s_0;z)}{F(s_0|s_0;z)(P_a-1)+z}.}
\\Finally the total rebinding probability of an in-contact pair on lattice is obtained by taking the limit $z\to 1$:
\eq{}{
H_{reb}=\lim_{z\to 1}H(s_0|s_1;z)=\frac{P_a}{P_a+\frac{1}{F(1)}-1},}
where $F(1)=F(s_0|s_0;z=1)\approx 0.256318$ (p. 153 in \cite{hughes1995}) is the return probability on HCP lattice. 

\subsubsection{Rebinding probability at long times}\label{arplt}
The asymptotic behavior of the rebinding-time probability distribution $H_n(s_0|s_1)$ at large $n$ can be estimated directly from the generating function. First we expand the generating function of the return probability $P_n(s|s)$ for the HCP lattice around $z=1$ up to the $O(1-z)$ term (see Eq. D.8b in \cite{montroll1965random} and Eq. A.237 in \cite{hughes1995})
\eq{}{P(s|s;z)\approx P(1)-c_1\sqrt{1-z}+O(1-z),}
where $P(1)=P(s|s;z=1)\approx 1.344661$ and $c_1=3^{3/2}/2\pi$.
\\The corresponding expansion of the generating function of $F_n(s|s)$ is then
\meq{expfz}{F(s|s;z)&=1-\frac{1}{P(s|s;z)}\\
&\approx 1-\frac{1}{P(1)-c_1\sqrt{1-z}}\\
&\approx 1-\frac{1}{P(1)}-\frac{c_1}{P(1)^2}\sqrt{1-z},}
where we have ignored the term equal to or higher than $O(1-z)$. 
\\Recall that the generating function of the rebinding-time probability distribution for the activation-limited case:
\meq{}{H(s_0|s_1;z)&=\frac{P_aF(s_0|s_0;z)}{z+F(s_0|s_0;z)(P_a-1)}\\
&=\frac{P_aF(s_0|s_0;z)}{z[1-F(s_0|s_0;z)(1-P_a)/z]}.}
By the expansion of the denominator we have 
\meq{}{H(s_0|s_1;z)&=\frac{P_aF(s_0|s_0;z)}{z}\left\{1+\frac{(1-P_a)F(s_0|s_0;z)}{z}+\left[\frac{(1-P_a)F(s_0|s_0;z)}{z}\right]^2+...\right\}\\
&=\frac{P_a}{z}\left\{F(s_0|s_0;z)+\frac{(1-P_a)}{z}F(s_0|s_0;z)^2+\left[\frac{(1-P_a)}{z}\right]^2F(s_0|s_0;z)^3+...\right\}.}
\\Substituting Eq. \eqref{expfz} into $H(s_0|s_1;z)$ and collecting the leading terms gives
\eq{}{H(s_0|s_1;z)\approx w\sqrt{1-z}+O(1-z),}
where
\meq{}{w&=-\frac{c_1P_a}{zP(1)^2}\left\{1-2\frac{(P_a-1)(P(1)-1)}{zP(1)}+3\left[\frac{(P_a-1)(P(1)-1)}{zP(1)}\right]^2+...\right\}\\
&=-\frac{c_1P_a}{zP(1)^2}\sum_{n=1}^\infty n (-1)^{n+1}\left[\frac{(P_a-1)(P(1)-1)}{zP(1)}\right]^{n-1}\\
&=-\frac{c_1P_a}{zP(1)^2}\left[1+\frac{(P_a-1)(P(1)-1)}{zP(1)}\right]^{-2}\\
&=-\frac{c_1P_a}{z\left[1+P_a(P(1)-1)+P(1)(z-1)\right]^2}\\
&=-\frac{c_1P_a}{\left[1+P_a(P(1)-1)\right]^2}.}
By means of singularity analysis of the generating function (see Eq. 2.3 of \cite{flajolet1990}), the corresponding asymptotic behavior of $H_n(s_0|s_1)$ as $n\to \infty$ is therefore 
\eq{asymp}{H_n(s_0|s_1)\approx -\frac{w}{2\sqrt{\pi}}n^{-3/2}+O(n^{-5/2}).}

\subsubsection{Rate coefficient at long times}\label{arclt}
From the definition of rate coefficient on lattice using the particle-pair formalism, we have the $m$-step reaction rate coefficient:
\meq{}{k_m&= k_a'\left[1-\sum_{n=0}^mH_n(s_0|s_1) \right], \ \text{for }m,n\in \mathbb{N},}\\
which can be rewritten as
\meq{}{k_m&=k_a'\left[1-\sum_{n=0}^\infty H_n(s_0|s_1) + \sum_{n=m}^\infty H_n(s_0|s_1) \right].
}
The first summation term is the total rebinding probability while the second term can be evaluated using the Euler-Maclaurin formula:
\meq{}{\sum_{n=m}^\infty H_n(s_0|s_1) &\approx \int _m^\infty dn \frac{w}{2\sqrt{\pi}}n^{-3/2}\\
&\approx \frac{w}{\sqrt{\pi n}}\\
&\approx \frac{lw}{\sqrt{6D\pi t}},}
where we have used the definition $nl^2=6Dt$ in the last step.\\
Now we have the asymptotic reaction rate as
\meq{}{\lim_{t\to\infty} k(t) &\approx k_a'\left[1-H_{reb}+\frac{lw}{\sqrt{6D\pi t}}\right].}
After rearrangement we have
\meq{}{\lim_{t\to\infty} k(t) &\approx k_a'(1-H_{reb})\left[1+\frac{lw}{(1-H_{reb})\sqrt{6D\pi t}}\right]\\
&\approx k_a'(1-H_{reb})\left[1+\frac{c_1P_al}{(1+(P(1)-1)P_a)\sqrt{6D\pi t}}\right].
}
Using the definition $k_a'(1-H_{reb})=k_{eff}'$, and applying the expressions for reaction acceptance probability in Eq. \eqref{PA} and voxel size in Eq. \eqref{vs}, we obtain the long-time approximation as
\eq{}{\lim_{t\to\infty} k(t) \approx k_{eff}'\left[1+\frac{k_aR}{(k_a+k_D)\sqrt{\pi Dt}}\right],}
which has the exact same form as the continuum case.

\subsection{Diffusion-influenced case \texorpdfstring{($k_a\gg k_D$, $\alpha<1$)}{Lg}.}\label{adic}
The derivation of the effective rate coefficient in the diffusion-influenced case differs from the activation-limited case due to the difference in the simulation scheme (see Algorithm 1 in main text), namely in the presence of non-unity step acceptance probability $P_w=\alpha$. The diffusion step $n$ is therefore no longer the same as the simulation step. Specifically, a successful arrival at a new target voxel (or a successful reaction attempt with a reactant) after $n=1$ step could have had multiple $k$ simulation steps in the past with hopping failures (or failed reaction attempts). As a result, the actual simulation time corresponding to $n$ steps is not a single value $nt'=nt_d\alpha$, but follows some distribution. 

The purpose of this section is to derive the long-time asymptotic behavior of the rate coefficient, which is independent of the transient time-dependent behavior. Hence, we parameterize the rebinding-time according to the eventful step $n$ (which will be incremented after a physical movement or a reaction attempt), rather than the actual simulation step $k$. The time-dependent behavior of rate coefficient on the other hand, will be treated in \ref{cont}.
 
As shown in the main text, the rebinding-time probability distribution $G_n(s_0|s_1)$ is defined as
\eq{g34}{G_{n+1}(s_0|s_1)=S_n(s_1|s_1)\ p(s_1\to s_0),\text{ for }n\in \mathbb{N}}
where $p(s_1\to s_0)$ is the reaction probability and $S_n(s_1|s_1)$ is the in-contact probability of a reactive pair after $n$ steps.

The reaction probability is defined as
\meq{rp35}{p(s_1\to s_0)&=P_a\alpha P_1(s_0|s_1)\sum_{k=0}^\infty \left\{[1-P_1(s_0|s_1)](1-P_w)\right\}^k\\
&=\frac{P_a\alpha P_1(s_0|s_1)}{1-(1-P_w)(1-P_1(s_0|s_1))},
} 
where the nominator term accounts for the probability of hopping to $s_0$ from $s_1$ and successfully reacting with the reactant located at $s_0$ in one diffusion step, while the denominator term comes from the infinite sum representing the total probability of unsuccessful escape to $s\in \{ \text{adjacent voxel of } s_1\} \setminus s_0$  at the previous simulation step\footnote{there are $k$ simulation steps in between each diffusion step $n$}. When $P_w=1$ and $\alpha=1$ as in the activation-limited case, then the reaction probability becomes $p(s_1\to s_0)=P_aP_1(s_0|s_1)$.

Next, we derive the generating functions of two first-passage time distributions $F_n(s_1|s_1)$ and $F_n(s_1|s_2)$ that correspond to the current scheme. We start from
\meq{34}{
F_{n+1}(s_1|s_1)&=\sum_s P_1(s|s_1)F_n(s_1|s),\ \text{for }n\in \mathbb{N}\\
&=P_1(s_0|s_1)\delta_{n,1}+P_1(s_1|s_1)\delta_{n,0}+P_1(s_2|s_1)F_n(s_1|s_2),
}
where the first term on the right-hand side relates to the failed reaction attempt $s_1\to s_0\to s_1$\footnote{only $s_1\to s_0$ is considered as a diffusion step, whereas the rejection $s_0\to s_1$ is not}, the second term describes the hop from $s_1\to s_1$, and the last term accounts for the trajectory $s_1\to s_2$, which is continued by a series of $n$ steps that have ended up in $s_1$ again.

From Eq. \eqref{34}, we obtain the generating function of $F_{n}(s_1|s_1)$ as
\eq{}{F(s_1|s_1;z)=z^2P_1(s_0|s_1)+zP_1(s_1|s_1)+zP_1(s_2|s_1)F(s_1|s_2;z).}
Thus we obtain
\eq{}{F(s_1|s_2;z)=\frac{F(s_1|s_1;z)-z^2P_1(s_0|s_1)-zP_1(s_1|s_1)}{zP_1(s_2|s_1)},}
where 
\eq{s1tos1}{F(s_1|s_1;z)=1-\frac{z^2P_1(s_0|s_1)}{P(s_0|s_0;z)-1}}
is given in terms of the generating function of $P_n(s_0|s_0)$
(the detailed derivation of Eq. \eqref{s1tos1} is given in \ref{fn11}).

Now, we define the probability that a particle is in-contact after $n$-step as:
\eq{sndl}{
S_n(s_1|s_1)=\gamma_1 S_{n-1}(s_1|s_1)+\sum^{n-1}_{m=0}\gamma_2 S_m(s_1|s_1)\ \bar F_{n-m-1}(s_1|s_2)+\delta_{n,0}S_0(s_1|s_1),\text{for }n\in \mathbb{N},
}
where the first term accounts for the trajectories $s_1\to s_0 \to s_1$ and $s_1\to s_1$, the second term represents the trajectories $s_1\to s_2\to s_1$ and the last term accounts for the initial condition. 
In detail, the coefficient
\meq{}{\gamma_1&=\left[(1-P_a\alpha)P_1(s_0|s_1)+P_w P_1(s_1|s_1)\right] \sum_{k=0}^\infty \left\{[1-P_1(s_0|s_1)](1-P_w)\right\}^k \\
&=\frac{(1-P_a\alpha)P_1(s_0|s_1)+P_w P_1(s_1|s_1)}{1-[1-P_1(s_0|s_1)](1-P_w)}, \\
}
accounts for the total probability of arrival at $s_1$ from a rejected reaction attempt (first sub-term) or from the adjacent neighbor $s_1$ (second sub-term) given that there was no successful escape to $s\in \{ \text{adjacent voxel of } s_1\} \setminus s_0$  at the last simulation step $k$ before the arrival, \\
while the coefficient
\meq{}{\gamma_2&=P_w P_1(s_2|s_1)\sum_{k=0}^\infty \left\{[1-P_1(s_0|s_1)](1-P_w)\right\}^k \\
&=\frac{P_w P_1(s_2|s_1)}{1-[1-P_1(s_0|s_1)](1-P_w)},}
accounts for the total probability of arriving at $s_2$ from $s_1$ given that there was no successful escape to $s\in \{ \text{adjacent voxel of } s_1\} \setminus s_0$  at the last simulation step $k$ before the arrival, \\
and finally $\bar F_n(s_1|s_2)=F_n(s_1|s_2)$ denotes the first-passage time distribution of the scheme with step-acceptance probability $P_w=\alpha$ (proof given in \ref{fn12}).

We then multiply Eq. \eqref{sndl} with $z^n$:
\eq{}{
S_n(s_1|s_1)\ z^n=\gamma_1 zS_{n-1}(s_1|s_1)\ z^{n-1}+\gamma_2 z\sum^{n-1}_{m=0}S_m(s_1|s_1)\ z^mF_{n-m-1}(s_1|s_2)z^{n-m-1}+\delta_{n,0}S_0(s_1|s_1)\ z^n,
}
and take the sum to infinity to obtain 
\eq{}{S(s_1|s_1;z)=\gamma_1 z S(s_1|s_1;z)+\gamma_2 z S(s_1|s_1;z)F(s_1|s_2;z)+S_0(s_1|s_1).}
\\After collecting the terms, we obtain the generating function of $S_n(s_1|s_1)$:
\eq{gfsp}{S(s_1|s_1;z)=\frac{S_0(s_1|s_1)}{1-\gamma_1 z-\gamma_2 z F(s_1|s_2;z)}.}
 
Substituting Eq. \eqref{rp35} and Eq. \eqref{gfsp} into Eq. \eqref{g34} then gives the rebinding-time probability distribution:
\eq{}{G_{n+1}(s_0|s_1)=\frac{P_a\alpha P_1(s_0|s_1)}{1-[1-P_1(s_0|s_1)](1-P_w)}S_n(s_1|s_1),\text{ for }n\in \mathbb{N},}
with the corresponding probability generating function
\meq{gf47}{G(s_0|s_1;z)&=\frac{P_a\alpha P_1(s_0|s_1)}{1-[1-P_1(s_0|s_1)](1-P_w)}S(s_1|s_1;z). }

In the diffusion-influenced scheme of Spatiocyte, we have $P_1(s_0|s_1)=1/12$, $P_1(s_1|s_1)=4/12$, $P_1(s_2|s_1)=7/12$ and $P_w=\alpha$. Using these parameters we then have the following quantities:
\meq{}{\gamma_1&=\frac{(1-P_a\alpha)+4\alpha}{12\left[1-11(1-\alpha)/12\right]}
,}
\meq{}{\gamma_2&=\frac{7\alpha}{12\left[1-11(1-\alpha)/12\right]}
,}
\eq{s2s1z}{F(s_1|s_2;z)=\frac{F(s_1|s_1;z)-z^2/12-4z/12}{7z/12},}
\meq{fz121}{
F(s_1|s_2;z=1)&=\frac{F(s_1|s_1;z=1)-1/12-4/12}{7/12}\\
&=\frac{8-1/F(1)}{7},}
where we have used definition Eq. \eqref{sums1s1} in Eq. \eqref{fz121}.\\
Using Eq. \eqref{fz121}, we obtain the limit of Eq. \eqref{gfsp} as:
\meq{szlim}{S(s_1|s_1;z=1)&=\left[1-\frac{(1-P_a\alpha)+4\alpha}{12\left[1-11(1-\alpha)/12\right]} -\frac{7\alpha}{12\left[1-11(1-\alpha)/12\right]} F(s_1|s_2;z=1) \right]^{-1}\\
&=\left[1-\frac{(1-P_a\alpha)+4\alpha}{12\left[1-11(1-\alpha)/12\right]} -\frac{7\alpha}{12\left[1-11(1-\alpha)/12\right]} \frac{8-1/F(1)}{7} \right]^{-1}\\
&=\frac{12\left[1-11(1-\alpha)/12\right]}{P_a\alpha-\alpha+\alpha/F(1)}.
}
Finally, we substitute Eq. \eqref{szlim} into Eq. \eqref{gf47} to obtain
\meq{}{
G(s_0|s_1;1)&=\frac{P_a\alpha/12}{1-11(1-\alpha)/12}\frac{12\left[1-11(1-\alpha)/12\right]}{P_a\alpha-\alpha+\alpha/F(1)}\\
&=\frac{P_a}{P_a+\frac{1}{F(1)}-1}
}
Therefore, we have the total rebinding probability as:
\eq{}{G_{reb}=G(s_0|s_1;z=1)=\frac{P_a}{P_a+\frac{1}{F(1)}-1}.}

\subsubsection{Return probability \texorpdfstring{$F_n(s_1|s_1)$}{Lg}} \label{fn11}

We denote $P_n(s| s_0)$ as the voxel occupation transition probability from $s_0$ to $s$. It is related to $F_n(s| s_0)$ via the convolution relation (\cite{hughes1995}, p. 121)
\meq{}{
P_n(s| s_0)=\delta_{ss_0}\delta_{n,0}+\sum_{j=1}^n F_j(s|s_0)P_{n-j}(s|s),\text{ for }n\in \mathbb{N}.
}
If a random walker started at $s_0$, it must go through $s_1$ before reaching the destination voxel $s$. Then we have
\eq{}{ P_n(s| s_0)=\delta_{ss_0}\delta_{n,0}+\delta_{s_0s_1}\delta_{n,1}P_1(s|s_1)+\sum_{j=1}^n F_j(s_1|s_0)P_{n-j}(s|s_1).}
Note that $P_n(s| s_1)=P_{n+1}(s|s).$
Thus, with $s_0=s_1$, we have
\eq{}{ P_{n+1}(s| s)=\delta_{n1}P_{2}(s|s)+\sum_{j=1}^n F_j(s_1|s_1)P_{n-j+1}(s|s).}
Multiplying both sides with $z^{n+1}$ gives
\eq{}{z^{n+1}P_{n+1}(s| s)=\delta_{n1}z^{n+1}P_{2}(s|s)+\sum_{j=1}^n z^{j}F_j(s_1|s_1)z^{n-j+1}P_{n-j+1}(s|s).}
Then taking the sum of both sides from $n=0$ to infinity gives
\eq{}{P(s|s;z)-P_0(s|s)=z^2P_2(s|s)+F(s_1|s_1;z)[P(s|s;z)-P_0(s|s)],}
where 
\eq{}{P(s|s;z)=\sum_{n=0}^\infty z^n P_n(s|s),\ F(s_1|s_1;z)=\sum_{n=0}^\infty z^n F_n(s_1|s_1)\text{ and } F_0(s_1|s_1)=0.}
\\As such, we have
\eq{s1s1}{F(s_1|s_1;z)=1-\frac{z^2 P_2(s|s)}{P(s|s;z)-1}.}
The total return probability to $s_1$ from $s_1$ is then
\meq{61}{
\sum_{n=0}^\infty F_n(s_1|s_1)&=\lim_{z\to {1^-}}F(s_1|s_1;z)\\
&=1-\frac{P_2(s_0|s_0)}{P(s_0|s_0;1^{-})-1}.
}
Using definition Eq. \eqref{FP} and $P_2(s_0|s_0)=1/12$, finally we have
\eq{sums1s1}{\sum_{n=0}^\infty F_n(s_1|s_1)=1-\frac{1/F(1)-1}{12}.}

\subsubsection{Return probability \texorpdfstring{$\bar F_n(s_1|s_2)$}{Lg}}\label{fn12}
If we increment the step count $n$ for every successful step to a new voxel, then the first-passage time distribution from $s_2$ to $s_1$ at step $n$ is given by
\eq{}{\bar F_n(s_1|s_2)=\sum^\infty_{m=n}\binom{m-1}{n-1}P_w^n(1-P_w)^{m-n}F_n(s_1|s_2),\text{ for }n\in \mathbb{Z}^+,}
where $P_w=\alpha$,is the step acceptance probability.
It can be shown that
\meq{bar12}{\bar F_n(s_1|s_2)&=P_w^nF_n(s_1|s_2)\sum^\infty_{m=n}\binom{m-1}{n-1}(1-P_w)^{m-n}\\
&=P_w^nF_n(s_1|s_2)\frac{1}{P_w^n}\\
&=F_n(s_1|s_2).}

\subsection{Continuous time limit of the diffusion-influenced scheme}\label{cont} 
In the diffusion-influenced scenario, Spatiocyte uses a different approach for hopping and reaction. Simulation progresses with a smaller time step $t'=t_d\alpha$ to resolve fast reaction events. We show that as $\alpha$ becomes smaller, the reaction and hopping events occur in a probabilistic manner that follows exponential time distribution. This property provides us with an approximation to study the time-dependent behavior of the reaction kinetics.
 
\subsubsection{Hopping time distribution}\label{htd}
Consider a single particle hopping on a completely vacant lattice. Let $P_w$ be the step acceptance probability for a particle heading to a vacant voxel. Then the probability of successful hopping after $m$ trials is
\eq{}{ P_h(t=m)=P_w(1-P_w)^{m-1}, \text{ for }m\in \mathbb{Z}^+,}
\\The survival probability (no hopping) until $m$th trial is then
\meq{}{
P_h(t>m)&=\sum_{m}^\infty P_w(1-P_w)^{m-1}\\
&=(1-P_w)^{m-1} .}
\\If we perform the trial every $\delta$ sec such that $P_w=\beta_1\delta$, where $\beta_1=t_d^{-1}$ is the average hopping rate per second.
The survival probability becomes
\meq{}{ 
P_h(t>m\delta)&=P_h(t>t')\\
&=(1-\beta_1\delta)^{m-1} ,}
where $t'=m\delta$. 
\\Similarly, we have
\meq{}{
P_h(t>t')&=(1-\beta_1\delta)^{\frac{t'}{\delta}-1}\\
&=\frac{(1-\beta_1\delta)^{\frac{t'}{\delta}}}{(1-\beta_1\delta)}.}
\\Taking the limit of small $\delta$, we then have
\meq{}{
\lim_{\delta\to 0}P_h(t>t')&=\left[\lim_{\delta\to 0}(1-\beta_1\delta)^{1/\delta}\right]^{t'}\\
&=\exp(-\beta_1 t').
}
Since $P_w=\alpha$, when $\alpha$ is small enough, the hopping time distribution of a particle approximates the exponential distribution 
\eq{psis}{\psi_h(t)=\exp(-\beta_1 t),}
with $\beta_1=t_d^{-1}$.

\subsubsection{Reaction time distribution}\label{artd}
Consider a reaction pair at an in-contact situation. The survival probability that they are still at the in-contact situation after $n$ steps is
\eq{}{S_n=(1-P_r-P_e)^n,\text{ for }n\in \mathbb{N},}
where $P_r=P_a\alpha/12$ is the reaction probability and $P_e=11P_w/12=11\alpha/12$ is the escape probability. 
Let the simulation trial performed at infinitesimal time $\delta$, such that $t'=n\delta=t_d\alpha$. \\
The survival probability as a function of time is then
\meq{}{
S(t')&=\lim_{\delta\to0}S_n\\
&=\lim_{\delta\to0} \left[1-\frac{\alpha}{12}(P_a+11) \right]^n\\
&=\lim_{\delta\to0} \left[1-\frac{\delta}{t_d}\frac{(P_a+11)}{12} \right]^{t'/\delta}\\
&=\left[\lim_{\delta\to0}\left(1-\beta\delta \right)^{1/\delta}\right]^{t'}\\
&=\exp(-\beta t').
}
where $\beta=({P_a+11})/{12t_d}$.
Note that the survival probability in this form includes both the probability of reaction and hopping events. Since the two events are independent of each other, the survival probability can be split into two separate terms:
\eq{}{S(t')=\exp(-\beta_1t')\exp\left(-\frac{11\beta_2t'}{12}\right),}
where $\beta_1=P_a/12t_d$ is the average reaction rate and $\beta_2=1/t_d$ is the average hopping rate. Therefore, the survival probability of the reaction also follows the exponential function 
\eq{psir}{\psi_r(t')=\exp(-\beta_1t').}

\subsubsection{Time dependent survival probability}\label{atdsp}

In summary, the survival probability of the reaction and hopping events are (from Eq. \eqref{psis} and Eq. \eqref{psir})
\meq{survival}{
\psi_r(t)&=\exp(-\beta_1 t)\ \text{, where}\ \beta_1=\frac{P_a}{12t_d},\\
\psi_h(t)&=\exp(-\beta_2 t)\ \text{, where}\ \beta_2=\frac{1}{t_d}.
}
Thus, the survival probability after one step is 
\eq{}{\psi(t)=\psi_r(t) \psi_h(t)=\exp\{-\beta t\},}
where $\beta=\beta_1+\beta_2$.
As a consequence, the survival probability of a reactive pair at short time $t$ after step $n$ follows the Poisson distribution:
\eq{sp}{ S_n(t)=\frac{(\beta t)^n}{n!}\exp(-\beta t),\text{ for }n\in \mathbb{N},}
where $S_0(t)=\exp(-\beta t)$.

\subsubsection{Rate coefficient at long times}\label{arcalt}
Here, we study the time-dependent kinetics of the diffusion-influenced scheme. We start with the definition of continuous rebinding-time probability density, and use it to express the time-dependent rate coefficient.
\\Denoting the continuous rebinding-time probability density after $(n+1)$ steps as
\eq{}{g_{n+1}(t)=\beta_1S_n(s_1|s_1;t),\text{ for }n\in \mathbb{N},}
where 
\eq{sdensity}{S_n(s_1|s_1;t)=\delta_{n,0} S_0(s_1|s_1;t)+\int_0^tdt'\sum_{j=0}^nS_{n-j}(s_1|s_1;t-t')F_j(s_1|s_1;t),}
is the survival time probability density of a particle that started and ended at $s_1$ on the $n$th step. The first term on the right-hand side of Eq. \eqref{sdensity} is the initial probability density $S_0(s_1|s_1;t)=\exp(-\beta t)$, while the last term involves two convolutions: the continuous time convolution and the discrete step convolution nested inside the time convolution.  $F_n(s_1|s_1;t)$ is the first-passage time density at the $n$th step, defined as
\meq{fptd}{
F_n(s_1|s_1;t)&=F_n(s_1|s_1)\left[ \vphantom{\int} \delta_{n,1}\beta_2\exp(-\beta t) \right.\\
&\left.+(1-\delta_{n,1})\int_0^tdt'\beta_2\exp(-\beta t')\frac{\beta_2^n (t-t')^{n-1}}{(n-1)!}\exp(-\beta_2(t-t')) \right],}
for $n\in \mathbb{Z}^+$. Intuitively, the first term describes the first-passage time distribution for single step while the second term accounts for the convolution of the probability of time required for the $n-1$ steps after the first step.

The continuous rebinding-time probability density is related to the rate coefficient of the particle-pair formalism through:
\eq{}{k(t)=k_a'\left[1- \int_0^t dt'g(t')\right],}
as shown in the main text.
We then take the Laplace transform of $k(t)$ which is easier to work with:
\eq{lpkt}{s\hat{k}(s)=k_a'[1-\hat{g}(s) ].} 
Note that $\hat{g}(s)$ is related to the rebinding-time and survival-time probability densities via:
\eq{gs84}{\hat{g}(s)=\sum_{n=1}^\infty \hat{g}_n(s_1|s_1;s)=\beta_1\sum_{n=1}^\infty\hat{S}_n(s_1|s_1;s).}
The corresponding Laplace transform of Eq. \eqref{sdensity} is given as
\eq{lps}{\hat{S}_n (s_1|s_1;s)=\frac{\delta_{n,0}}{s+\beta}+\sum_{j=0}^n\hat{S}_{n-j}(s_1|s_1;s)\hat{F}_j(s_1|s_1;s),}
where
\eq{lpf}{\hat{F}_n(s_1|s_1;s)=F_n(s_1|s_1)\frac{\beta_2}{s+\beta}\left[\delta_{n,1}+(1-\delta_{n,1})\left(\frac{\beta_2}{s+\beta_2}\right)^n \right].}
The infinite sum of Eqs.\eqref{lps} and \eqref{lpf} are 
\meq{}{\sum_{n=0}^\infty \hat{S}_n(s_1|s_1;s)&=\frac{1}{s+\beta}+\sum_{n=0}^\infty \sum_{j=0}^n \hat{S}_{n-j}(s_1|s_1;s)\hat{F}_j(s_1|s_1;s), \\
&=\frac{1}{s+\beta}+\sum_{n=0}^\infty \hat{S}_{n}(s_1|s_1;s) \sum_{n=0}^\infty \hat{F}_n(s_1|s_1;s), \\
&=\frac{1}{s+\beta}\left[1-\sum_{n=0}^\infty \hat{F}_n(s_1|s_1;s) \right]^{-1},\\
}
\meq{}{\sum_{n=0}^\infty \hat{F}_n(s_1|s_1;s)&=\frac{\beta_2}{s+\beta}\left[F_1(s_1|s_1)+\sum_{n=2}^\infty F_n(s_1|s_1) \left(\frac{\beta_2}{s+\beta_2}\right)^n \right]\\
&=\frac{\beta_2}{s+\beta}\left[F_1(s_1|s_1)\frac{s}{s+\beta_2}+\sum_{n=0}^\infty F_n(s_1|s_1) \left(\frac{\beta_2}{s+\beta_2}\right)^n \right]\\
&=\frac{\beta_2}{s+\beta}\left[F_1(s_1|s_1)\frac{s}{s+\beta_2}+F\left(s_1|s_1;z=\frac{\beta_2}{s+\beta_2}\right) \right],}
where $F(s_1|s_1;z)$ is the generating function, $\sum_{n=0}^\infty F(s_1|s_1)z^n$ is as defined in \eqref{s1s1}.
\\Hence, we have
\eq{sumss}{\sum_{n=0}^\infty \hat{S}_n(s_1|s_1;s)=[s+\beta-sF_1(s_1|s_1)z-\beta_2F(s_1|s_1;z)]^{-1},}
where $z=\beta_2/(s+\beta_2).$

Substituting Eq. \eqref{sumss} into Eq. \eqref{gs84} and by the final value theorem, we obtain the long-time behavior of $k(t)$ by taking the limit $s\to 0$ in Eq. \eqref{lpkt}. 
\\Assuming the asymptotic Laplace form of the rate coefficient on lattice (\cite{agmon1990} Eq. 2.37a):
\eq{lpks}{s\hat{k}(s)\approx k_{eff}'(1+a_{eff}'\sqrt{s/D}+...).}
We then set $s=0$ to obtain the effective lattice reaction rate constant: 
\meq{}{k_{eff}'&=k_a'[1-\hat{g}(0)].}\\
Evaluating $\hat{g}(0)$ by referring to Eq. \eqref{gs84}, we then get
\meq{}{k_{eff}'&=\frac{3\sqrt{2}P_aDl}{1+P_a/(1/F(1)-1)},\\
}
which is consistent with the result shown in main text.
\\The second order term of Eq. \eqref{lpks} is evaluated by expanding $s\hat{k}(s)$ around $s=0$:
\meq{}{\lim_{s\to0} \frac{d}{d\sqrt{s}}s\hat{k}(s)&=\lim_{s\to0} \frac{d}{d\sqrt{s}}(-k_a'\hat{g}(s))\\
&=\frac{-k_a'\beta_1}{\sqrt{\beta_2}}\lim_{q\to0}\frac{d}{dq}\sum_{n=0}^\infty \hat{S}_n(s_1|s_1;s)\\
&=k_a'\beta_1\sqrt{\beta_2}\left[\beta-\beta_2F(1)\right]^{-2}\lim_{q\to0}\frac{d}{dq}F(q)\\
&=\frac{2k_a'P_a}{\sqrt{\beta_2}}\left[1+\frac{P_a}{1/F(1)-1}\right]^{-2}\lim_{q\to0}q\frac{d}{dz}P(s_0|s_0;z),\\
}
where $q=\sqrt{s/\beta_2}$ and $z=1/(q^2+1)$.
\\Thus, by comparing the terms we obtain
\meq{}{
a_{eff}'&=\frac{\sqrt{D}}{k_{eff}'}\frac{2k_a'P_ab}{\sqrt{\beta_2}}[P_a\{P(s_0|s_0;1)-1\}+1]^{-2}\\
&=\frac{\sqrt{2/3}bP_al}{1+P_a/(1/F(1)-1)},}
where 
\eq{}{
b=\lim_{q\to0}q\frac{d}{dz}P(s_0|s_0;z)=\frac{3\sqrt{3}}{4\pi}.}
\\Applying the definitions of reaction acceptance probability \eqref{PA} and voxel size \eqref{vs}, we obtain
\eq{}{a_{eff}'=\frac{k_aR}{k_a+k_D}.} 
Note that the corresponding time domain form of Eq. \eqref{lpks} is given as
\eq{}{k(t)\approx k_{eff}'\left[1+a'_{eff}\sqrt{\pi Dt}+...\right ].}
Hence, the long-time behavior of the lattice rate coefficient follows the same form as in the continuum case:
\eq{}{k(t)\approx k_{eff}'\left[1+\frac{k_aR}{(k_a+k_D)\sqrt{\pi Dt}}\right].}

\section{Production-degradation process}\label{apdp}
In the coupled reactions  $\emptyset {\stackrel{k_1}{\xrightarrow{}}}A$, \ $A+B{\stackrel{k_2}{\xrightarrow{}}}B$, the survival probability of a newly produced $A$ molecule in an equilibrated pool of $B$ is
\eq{}{S_{rad}(t|eq)=\exp \left[-[B]\int_0^t k_{rad}(t')dt' \right].}
where $[B]$ is the concentration of $B$ and $k_{rad}(t)$ is the irreversible rate coefficient according to the radiation boundary condition. Since $A$ is removed from the system via the bimolecular reaction, the concentration of $A$ will eventually reach a steady-state. The corresponding mean lifetime of the decay $\tau$ is used to define the steady-state rate coefficient $k_{ss}$ \cite{agmon1990}:
\eq{eqkss}{\tau=\frac{1}{[B]k_{ss}}\equiv\int_0^\infty S_{rad}(t|eq)=\hat{S}_{rad}(0|eq),}
where the hat denotes Laplace transform.
\\For small $[B]$, $k_{ss}$ is given by (Eq. 4.5 in \cite{agmon1990})
\eq{}{k_{ss}\approx k_{on}\left\{1+\left[4\pi(R_{eff}^{rad})^3 [B]\right]^{1/2} \right\},}
where $k_{on}=4\pi DR_{eff}^{rad}$ is the macroscopic rate constant, $R_{eff}^{rad}=k_a R/(k_a+4\pi RD)$ is the effective radius and $k_a$ is the intrinsic reaction rate constant.
\\The equilibrium concentration of $A$ is then
\eq{ksseq}{[A]=\frac{k_{1}}{k_{ss}[B]}.}

\end{appendices}

\end{document}